# Tuning the catalytic activity of graphene nanosheets for oxygen reduction reaction via size and thickness reduction†

*John Benson[a], Qian Xu[b], Peng Wang[b], Yuting Shen[c], Litao Sun[c], Tanyuan Wang[d], Meixian Li[d] and Pagona Papakonstantinou*[a]*

[a] Engineering Research Institute, School of Engineering, University of Ulster, Newtownabbey BT37 0QB, UK

[b] National Laboratory of Solid State Microstructures and College of Engineering and Applied Sciences, Nanjing University, 22 Hankou Road, Gulou, Nanjing, 210093, P. R. China

[c] SEU-FEI Nano-Pico Center, Key Laboratory of MEMS of Ministry of Education, Southeast University, Sipailou 2, Nanjing 210096, P. R. China

[d] College of Chemistry and Molecular Engineering, Peking University, Beijing 100871, P.R.China.

† Electronic Supplementary Information (ESI) available: [Methods, instrumentation and supplementary Figures].

**ABSTRACT**

Currently, the fundamental factors that control the oxygen reduction reaction (ORR) activity of graphene itself, in particular the dependence of the ORR activity on the number of exposed edge sites remain elusive, mainly due to limited synthesis routes of achieving small size graphene. In this work, the synthesis of low oxygen content (< 2.5±0.2 at %), few layer graphene nanosheets with lateral dimensions smaller than a few hundred nm was achieved using a combination of ionic liquid assisted grinding of high purity graphite coupled with sequential centrifugation. We show for the first time, that the graphene nanosheets possessing a plethora of edges exhibited considerably higher electron transfer numbers compared to the thicker graphene nanoplatelets. This enhanced ORR activity was accomplished by successfully exploiting the plethora of edges of the nanosized graphene as well as the efficient electron communication between the active edge sites and the electrode substrate. The graphene nanosheets were characterized by an onset potential of -0.13 V vs. Ag/AgCl and a current density of -3.85 mA/cm$^2$ at -1 V, which represent the best ORR performance ever achieved from an undoped carbon based catalyst. This work demonstrates how low oxygen content nanosized graphene synthesized by a simple route can considerably impact the ORR catalytic activity and hence it is of significance in designing and optimizing advanced metal-free ORR electrocatalysts.

**KEYWORDS**: graphene nanosheets, oxygen reduction reaction, electrocatalyst, edges, ionic liquid exfoliation.

**Introduction**

Increasing demands for clean energy have stimulated extensive research on the development of technologies that can effectively convert chemical energy into electricity with high efficiency and at low cost. Catalysts for oxygen reduction reaction (ORR) are at the heart of key electrochemical technologies including low temperature polymer electrolyte membrane (PEM) fuel cells and metal air batteries[1-5]. The existing catalysts of Pt group metals are highly efficient but too expensive and rare to be useful for mass production[6-8]. Graphene promises a low-cost alternative to the precious metals[9-12]. In particular there has been an explosion of studies on introducing various heteroatoms (e.g. N, B, P, S, and I)[13-16] or a combination of those[17, 18] into graphene demonstrating a competitive ORR activity compared to the benchmark Pt catalyst. The phenomenon has been rationalized using density functional calculations, where it was found that the heteroatom induces an uneven charge distribution in the adjacent sites, which alters the local spin or charge density. This promotes O adsorption and facilitates efficient ORR performance[19, 20].

However, despite the remarkable progress, fundamental factors that control the ORR activity of graphene itself, in particular the dependence of the ORR activity on the number of exposed edge sites remains elusive due to limited synthesis routes of achieving small size and preferably oxygen free graphene sheets. For large graphene sheets an altered electronic structure is expected at the basal plane as compared to the edge region due to symmetry breaking of the honeycomb lattice. Graphene has two main kinds of edge terminations according to their shape, named zigzag and armchair edges. These two edges have different electronic structures. According to theoretical and experimental studies[21, 22] a π electron state called *edge state* is created along the zigzag edges, whereas no such state is present at the armchair edges. This characteristic *edge* *state*, which has a large local density of states and is spin polarized, gives rise to electronic, magnetic and chemical activities in the zigzag edges of graphene. Recent experimental studies on large area graphene have provided supporting evidence for the high ORR activity of the edges[23]. Hence, it can be insinuated that a prerequisite for a highly ORR electroactive graphene catalyst is the ability to nanostructure graphene in a manner conducive of maximizing the number of exposed edge sites (preferably zigzag) relative to catalytically inert basal planes sites [24-26].

As far as we are aware, no attempt has been made to synthesize graphene sheets, simultaneously possessing low oxygen content and small lateral dimensions with the view of examining their electroactivity. It is worth noting that under ambient conditions a low concentration of oxygen (< 3 at % depending on the amount of edges) will always be present, since open graphene edges can be easily terminated by oxygen groups. Carbon materials possessing oxygen species are known to favor the 2e- process in ORR[27]. Moreover, progressive addition of oxygen content in carbon materials has been shown to limit and slow down the electron transfer reactions[28]. Therefore, introducing purposely more oxygen in the graphene is not considered to have a beneficial effect for achieving efficient (towards 4e-) ORR.

Clearly, there remains a dearth of experimental data on the ORR activity of low oxygen level graphene nanosheets. Here we have developed a novel approach for the production of low oxygen content few layer graphene nanosheets via simple and green ionic liquid assisted mechanical exfoliation of graphite[29] combined with gradient centrifugation steps. The grinding of graphite with a small quantity of ionic liquid produces a gel due to the π-π interactions between the graphite and ionic liquid. During the grinding process the ionic liquid acts as a lubricant, allowing the breaking of graphite platelets to smaller sizes and at the same time helps exfoliation of graphite layers through shear forces exerted on the graphite flake. Although the bucky gels produced by grinding of carbon nanomaterials with ionic liquid are known since 2003[30] and they have found use in a number of applications such as supercapacitors, biosensors and actuators[31], the production of nanosized graphene with low amount of oxygen has not been reported so far to our knowledge. Using our process, defect formation on the crystalline plane of graphene, or chemical reactions due to mechanochemical effects are avoided resulting in high quality material. The edges as well as the basal plane of the graphene nanosheets are free from any additional functional groups, possessing only a small amount of oxygen (< 2.5±0.2 at %) mainly inherited from the starting graphite and exposure of edges to atmosphere.

The sequential centrifugation steps employed here help to isolate graphene nanosheets with small lateral dimensions. The process dramatically enhances the presence of small size graphene nanosheets, thus providing abundant catalytic sites. We have taken extensive measures to ensure the grinding method does not involve metal components unlike many ball milling techniques, and therefore metal contamination from the grinding instrument is avoided[32, 33]. As a result this new synthesis approach is ideal for revealing and gaining more knowledge on the role of graphene edges for the ORR.

We show for the first time, that the ORR activity of graphene nanosheets possessing a plethora of edges with limited amount of oxygen (< 2.5±0.2 at %) is substantially improved when compared to thicker graphene nanoplatelets as revealed by enhanced electron transfer numbers. Therefore the current study of the ORR on low oxygen content graphene nanosheets is expected to provide new insight into the design and feasible synthesis of more advanced graphene-based catalysts.

**Results and Discussion**

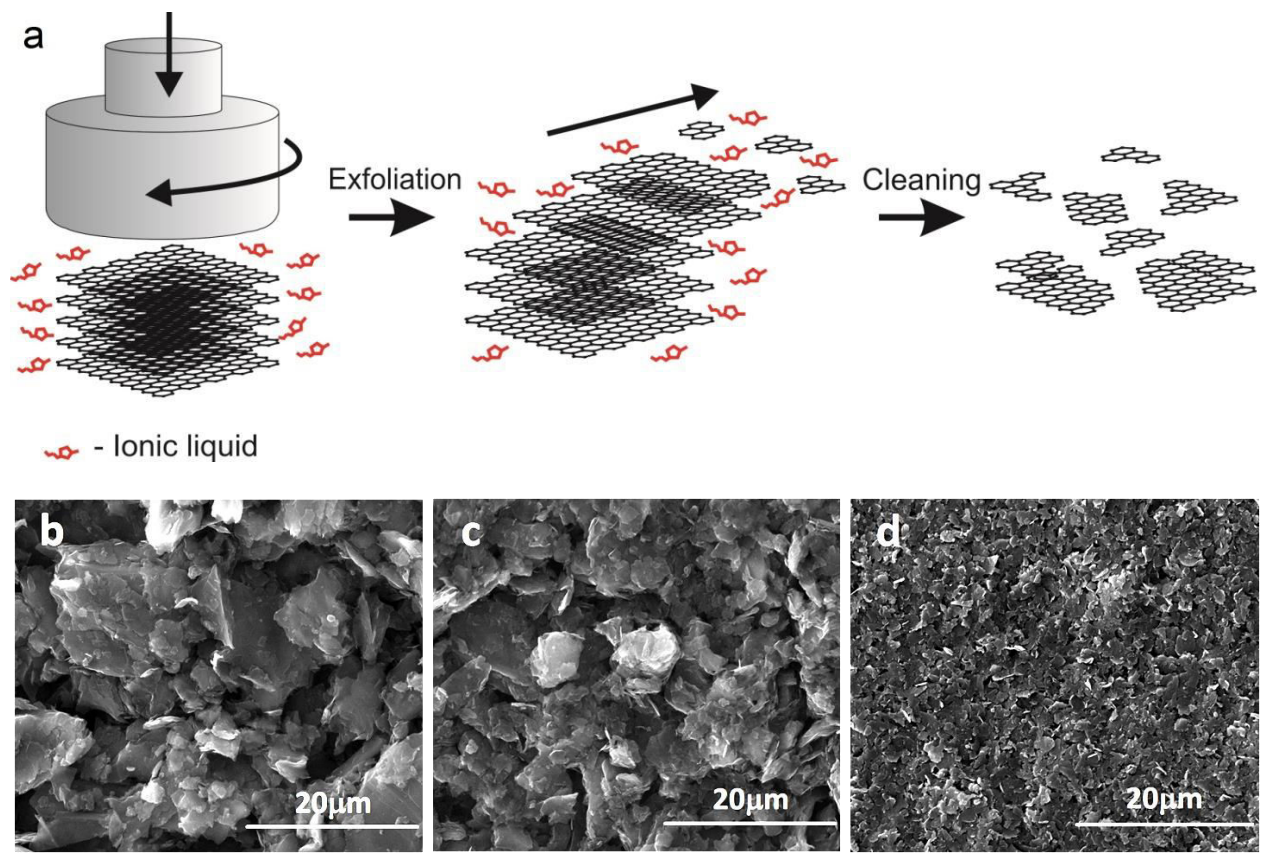

**Figure 1.** a) Schematic representation of exfoliation of graphite. Shear forces exfoliate layers in a mechanical mortar grinder with ionic liquid, resulting in cleaned graphene nanosheets, free from functional groups. SEM images of b) pristine graphite; sediment of ground material obtained at successive centrifugation speeds of: c) 1,000 rpm (1K) and d) 10,000 rpm (10K).

In a typical experiment, grinding was carried out in a planetary grinding machine in the presence of high purity graphite ($\geq$99.99%) and ionic liquid (IL, 1-Butyl-3- methylimidazolium hexafluorophosphate, BMIMPF$_6$, $\geq$97.0%) as depicted in the schematic of Figure 1. After grinding, the produced gel was washed in a mixture of DMF and acetone to remove the ionic liquid[34]. The supernatant from a DMF solution that contained the graphene nanosheets was collected through sequential centrifugation at 1,000, 3,000 and 10,000 rpm speeds[35]; for more information see supporting information. In this work the products are named as XK for convenience, where XK denotes the centrifugation speed in thousands of rpm used to collect the sediment. The pristine graphite consists of large graphite platelets (~45 μm) as well as smaller irregular graphite flakes. SEM images of pristine graphite and graphene nanosheets produced through various centrifugation speeds at both low- and high-magnification can be seen in Figure 1 and S1. Grinding in ionic liquid coupled with sequential centrifugation gave rise to progressive size selection. Centrifugation at 10K resulted in a dramatic decrease in sheet size (< 1 μm), (Figure 1S), exposing a large fraction of edge sites compared to lower centrifugation speeds.

High-resolution TEM (HRTEM) studies provided information on the crystalline quality of the sheets as well as their thickness. Characterization of the residue products from the sequential centrifugation steps by TEM ensured that the final 10K product consists of a few-layer sheets (up to 9) with overall lateral dimensions smaller than a few hundred nm in contrast to 1K and 3K, which comprised of thicker and larger size products, (Figure 2a-i and Figure S2-S6). In fact for the 10K product, small size sheets are self-assembled onto the surface of larger few layer graphene sheets, most probably due to pi-pi interactions. Such self-assembled configuration allows more efficient transport. It is apparent that the nanosheets have a flat morphology, different to that reported[36] for reduced graphene oxide with the partially crumbled nature that originates from the defective structure formed during the fabrication of graphene oxide.

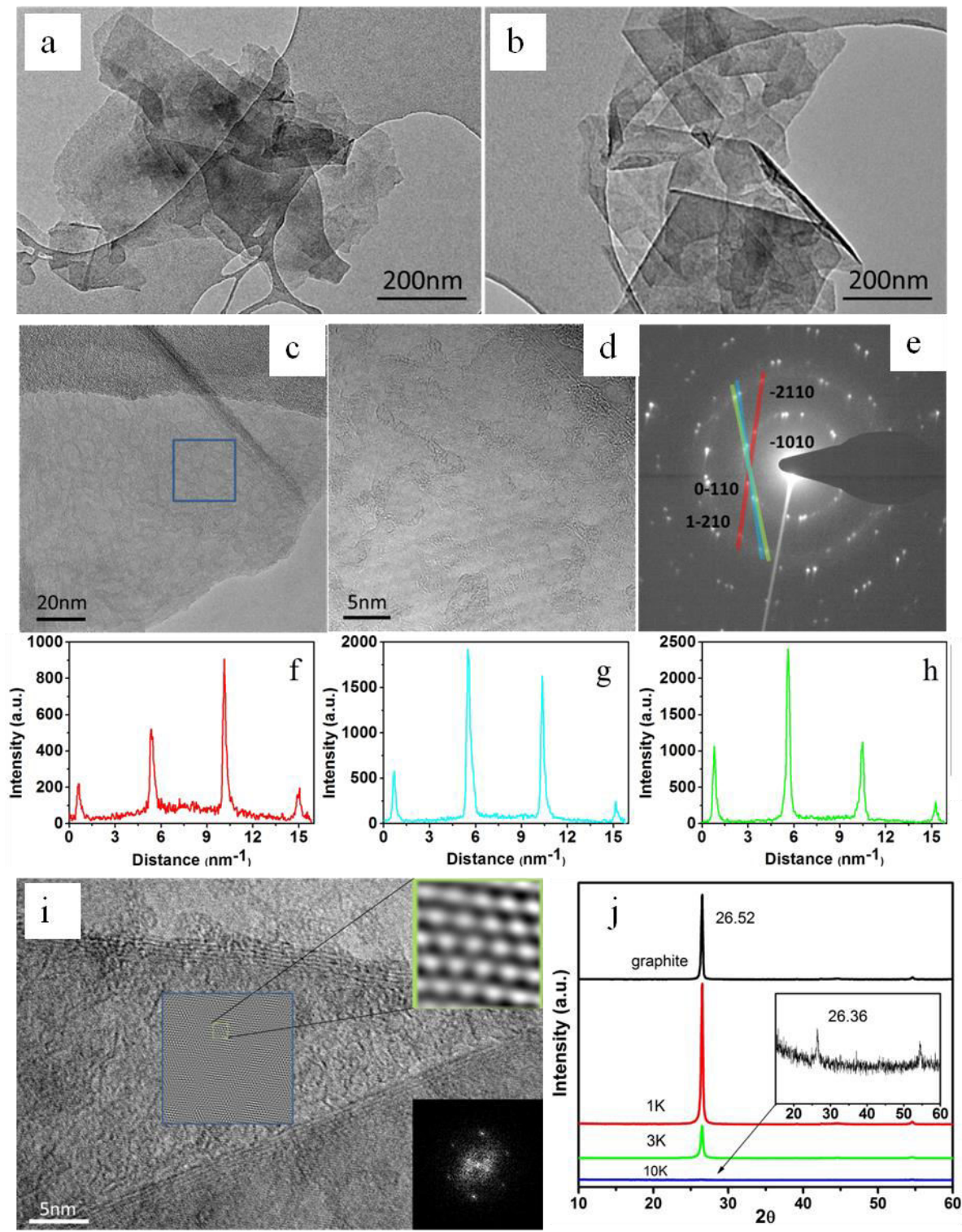

**Figure 2.** TEM images of typical products found in the 10K sediment. a), b) The 10K product is dominated by nanosized few layer graphene assembled on larger sheets. c) Low magnification

TEM images from 10K product acquired on an aberration correction Titan 80-300 operated at 80 KV, with d) magnified region of the image in c). e) is the corresponding electron diffraction pattern of d) which indicates that there are at least three different orientations of graphene. f - h) Intensity profile of diffracted spot extracted along the -2110 to 1-210 axis for three oriented patterns in e). Ratios of the intensity of {1100} and {2110} are all >1, which indicates that the graphene in each orientation is formed with a single layer. i) HR-TEM of a 4-5 layer graphene: (Top right corner) an enlargement of the noise-filtered area as marked with a blue square, where the clearly visible parallel lines demonstrate the regular period of the lattice planes; (Bottom right corner) a Fourier transform of the blue marked area of the raw image. j) XRD diffraction patterns graphite together with those of 1K, 3K and 10K products. Inset shows magnified section of 10K product. Peak intensity (002) is 0.5% of pristine graphite indicating a high degree of exfoliation and a decrease in lateral dimensions.

HRTEM image performed on one typical cross-section edge of the nanosheet, indicates an interlayer spacing of 0.326 nm for (001) plane, (Figure S5). Selected area electron diffraction pattern from a 10K nanosheet about 3-4 layers reveals at least 3 sets of hexagonal patterns, indicating different orientations. Computational and experimental studies[37-39] have shown that TEM selected area electron diffraction (SAED) can convincingly distinguish monolayer from multilayer graphene by comparing the intensity ratio of first-order to second-order diffraction spots and diffraction peak intensity variation with respect to the tilt angle of the sample holder. It was observed that multilayer graphene with Bernal (AB) stacking gives higher diffracted intensities for the outer {2110} type spots than inner [1100]-type spots ( I{1100}/I{2100}<1), because of interference between electrons scattered from the A-type and B-type layers. Conversely, for a single graphene layer the intensities of the inner {1100}-type spots are stronger than those of the {2100}-type spots. For our materials line profiles extracted along the -2110 to 1-210 axis for the three oriented diffraction patterns were used to identify the thickness of layers along these directions[37, 38]. The ratios of {1100} to {2110} for all 3 different orientations are all larger than 1, indicating that the stacking consists of individual graphene layers along each of the three different directions.

A clear hexagonal arrangement could be seen in the HRTEM images of Figure 2i revealing that the exfoliated graphene sheets are of high crystalline quality and at the same time confirming that ionic liquid assisted grinding did not damage or disrupt the hexagonal structure of graphene. In Figure 2i, a portion of the image has been digitally noise filtered. The clearly visible parallel lines demonstrate the regular period of the lattice planes. EDAX of 10K product (Figure S6) confirms the presence of only carbon and oxygen, indicating the product is of high quality with absence of impurities.

The morphology of the 3K and 10K catalysts was further confirmed with AFM measurements. Figure S7 is an AFM image of 3K product, where clearly distinguishable layers of both small (400 nm) and larger (~1 μm) lateral dimensions are observed with a variable sheet thicknesses between 2-6 nm. Graphene sheets less than 10 layers thick with lateral diameters of a few hundred nm were observed for 10K catalyst (Figure S8).

The progressive exfoliation and reduction in lateral size of the centrifugation products was corroborated by XRD measurements. As can be seen in Figure 2j the pristine graphite exhibited a typical strong peak at 26.52°, corresponding to an interlayer d-spacing of 0.3358 nm of the (002) plane. The centrifugation products displayed a progressive decrease in peak intensity. The 10K product retained only 0.5% of the (002) peak intensity of the pristine graphite flakes and exhibited a dramatic broadening (Table S1), indicating considerably reduced lateral size[40-42]. The slight increase in interplanar spacing (from 0.3358 nm to 0.3378 nm) is probably due to the fact that the layers although they are aggregated during film formation, they are decoupled from each other.

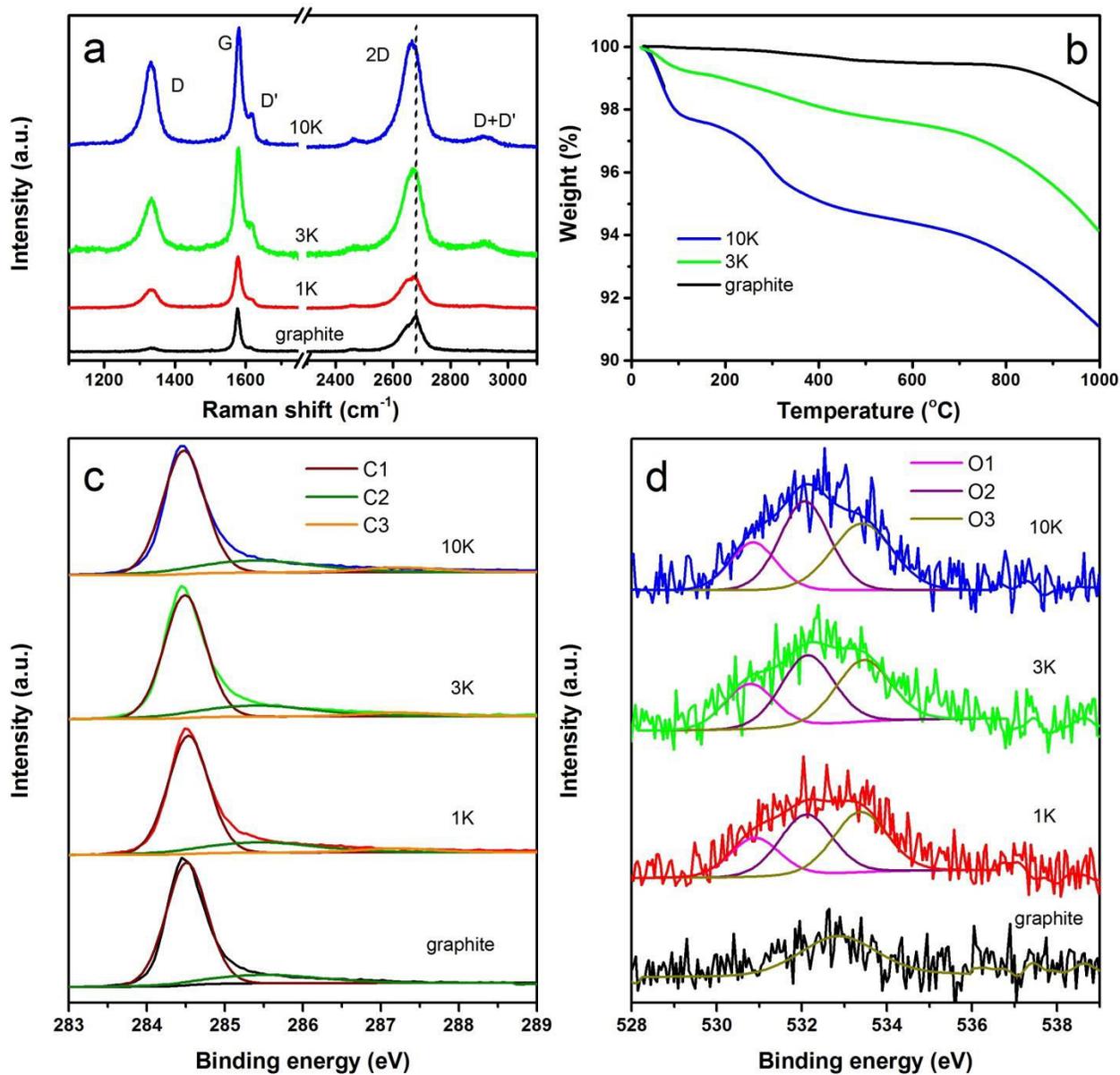

**Figure 3.** a) Raman spectra showing increased D band intensity and a 2D peak downshift with increased symmetry indicating exfoliation and flake size reduction. b) TGA plots of graphite, 3K and 10K. Measurements were carried out under $N_2$ atmosphere with a heating rate of 5°C min

over the temperature range 24°C-1000°C. High resolution c) C1s and d) O 1s deconvoluted XPS spectra of graphite, 1K, 3K, and 10k.

To further evaluate the structure of the graphene nanosheets, Raman spectra were taken on a Labram Raman spectroscope with a laser excitation wavelength of 632.8 nm and a beam size of approximately 2 μm. Raman spectroscopy is a powerful tool for identifying the number of layers and structural defects of graphene[43, 44]. Figure 3a shows the evolution of the spectral features of the products upon various centrifugation cycles together with that of graphite. Pristine graphite can be seen to display the characteristic Raman 2D signal of graphite, having a double peak structure. It exhibits a small detectable D band at around 1334 cm$^{-1}$, and a D-band to G-band intensity ratio ($I_D/I_G$) of 0.11. The presence of the small D band, which is a defect- related peak, most probably is linked with the presence of small flakes, which coexist with the large graphite platelets as revealed by SEM images, (Figure 1 and S1). As expected, the ground products produced from the sequential centrifugation showed strong D bands around 1332 cm$^{-1}$ with the $I_D/I_G$ ratios in the range of 0.37-0.72, indicating progressive size reduction, which is consistent with the TEM and SEM results (Figure 1-2 and Figure S1–S5). The progressive increase of the D peak suggests that the D-band should arise predominantly from edge defects upon increased centrifugation speeds, rather than basal plane defects, since the nanosheet lateral sizes were typically smaller than the laser spot diameter (2 μm)[44, 45]. The observed $I_D/I_G$ values are lower than those of large size reduced graphene oxide, where a considerable amount of disorder remains after the reduction of graphene oxide material[46-48]. The disorder related peaks, D′ and D + D′ became apparent at 3K and 10K products. The 2D band becomes more symmetrical with increasing the centrifugation speed and is downshifted from ~2675 cm$^{-1}$ at 1K to ~2666 cm$^{-1}$ at 10K. The development of a more symmetrical 2D line shape together with the blue shift of 2D band relative to graphite, suggests that the 10K product consists of few-layers[44]. The results confirm that the deposited nanosheets consist of a disordered arrangement of randomly restacked but weakly interacting few-layer graphene nanosheets in agreement with the TEM observations.

The smaller size and exfoliation under progressive centrifugation steps was also confirmed by Thermogravimetric analysis (TGA. TGA graphs (Figure 3b) of both 3K and 10K products are presented and compared with that of intact graphite. The latter is thermally stable when heated up to 1000 °C under nitrogen atmosphere, indicated by a weight loss of only 1.90%; however, the 3K and 10K products show a gradual higher weight loss of 5.8 and 8.9% respectively reflecting the smaller lateral dimensions and higher degree of exfoliation. It is expected that a few layer graphene with small dimensions will have lower combustion temperature compared to large area graphitic sheets[49]. The 10K product exhibits a higher weight loss than the rest of the samples due to the presence of a large amount of a few layer nanosized graphene. The graphite exhibits the greatest overall thermal stability due to the extended stacking and large lateral size of graphite platelets and as a result, requires a much higher temperature for decomposition. Our findings are in agreement with the report by Kinoshita et al.[50], which indicates that thermal decomposition temperatures for natural graphite powders decrease as the density of edge plane sites increase.

Differential Scanning Calorimetry (DSC) DSC-TGA plots (Figure S9) show the 10 K product typically exhibited two well-defined mass losses, which took place at 25-125 °C and 140–400 °C, with peaks at 62.6°C and 292°C, respectively. These observed weight losses can be attributed to gasification of adsorbents and entrapped DMF between the graphene layers (boiling point 153 °C) as well as the thermal decomposition of oxygen edge groups.

XPS analysis provided explicit information on the nature and exact content of any functional groups brought by the exfoliation. Figures 3c-d and Tables S2–S4 summarize the results of the XPS investigation of graphite and centrifugation products. XPS wide spectra from all samples show a strong C1s peak at 284.5 eV associated with the sp$^2$ hybridized framework and a minor O1s peak at approximately 532.5 eV (Figure S10a). The high resolution carbon C1s (Figures 3c), spectra were fit with up to three peaks using binding energies of 284.47, 285.40, and 287.72 eV. These peaks correspond to the following carbon components: sp$^2$ (C1), C–O associated with ether and phenolic groups (C2) and C=O (C3)[51, 52] groups respectively. The N1s, F1s, P2p, Co2p, Co2p, Fe2p, Ni2p, Mn2p core level spectra show no observable nitrogen, fluorine, phosphorous and metal traces, (Figure S10b-h). These results confirm the absence of any functional groups or metals accrued from the ionic liquid, DMF solvents or grinder. Thus the ionic liquid used during the grinding process does not bind to the inert basal graphene plane and is effectively removed to a level below the XPS detection limit. Prolonged sonication in DMF has also been avoided and therefore nitrogen moieties cannot be inherited by DMF[53]. They also confirm that the grinding process does not introduce any traces of metal impurities in the composition of the products. The O1s spectra contain contributions from O1: Oxygen doubly bonded to aromatic Carbon (C=O) at 530.77 eV; O2: singly bonded oxygen to C–OH and C–O–C at 532.00eV and O3: singly-bonded oxygen O=C–OH, O=C–OR at 533.11 eV[51, 52]. An increase of no more than 1.5 at % was observed in the oxygen content of the ground centrifugation products compared to that of the starting graphite (starting graphite: 1.08 ±0.28 at %; 1K: 2.34±0.26 at%; 3K: 2.08±0.50 at%; 10K: 2.58±0.21 at %). This increase is due to the existence of a larger amount of edges, which can be attacked by oxygen in air. Figure S11 shows overlap of graphite and 10K C 1s are identical.

Using inductively coupled plasma mass spectrometry (ICP-MS) technique, we assessed the starting metallic-impurity content of natural graphite as well as the variation of the impurities in the centrifugation products 3K and 10K. Table 1, summarizes the results of the analysis.

|    | Graphite | 3K | 10K |
|----|----------|-----|-----|
| **Mn** | 0.66 | 0.28 | 0.14 |
| **Co** | 0.25 | 0.28 | 0.39 |
| **Ni** | 0.57 | 0.65 | 1.68 |
| **Fe** | <Detection limit | <Detection limit | <Detection limit |
| **Mo** | 0.45 | 0.42 | 0.03 |
| **P** | 6.62 | 65.40 | 70.60 |

**Table 1**. Impurity content (ppm) in starting graphite, 3K and 10K catalysts as determined by ICP-MS analysis.

It can be clearly seen from Table 1 that for graphite, 3K and 10K products, metal impurities (Mn, Co, Ni, Fe, Mo) are present in concentrations below 2 ppm, which clearly demonstrates that the starting graphite was of high purity and the grinding process did not introduced any metals contaminants to the 3K and 10K products. A small increase in the Phosphorous level from 6.62 ppm in the starting graphite to 70.60 ppm (0.0071 at %) in 10K, was observed, which originates from the ionic liquid.

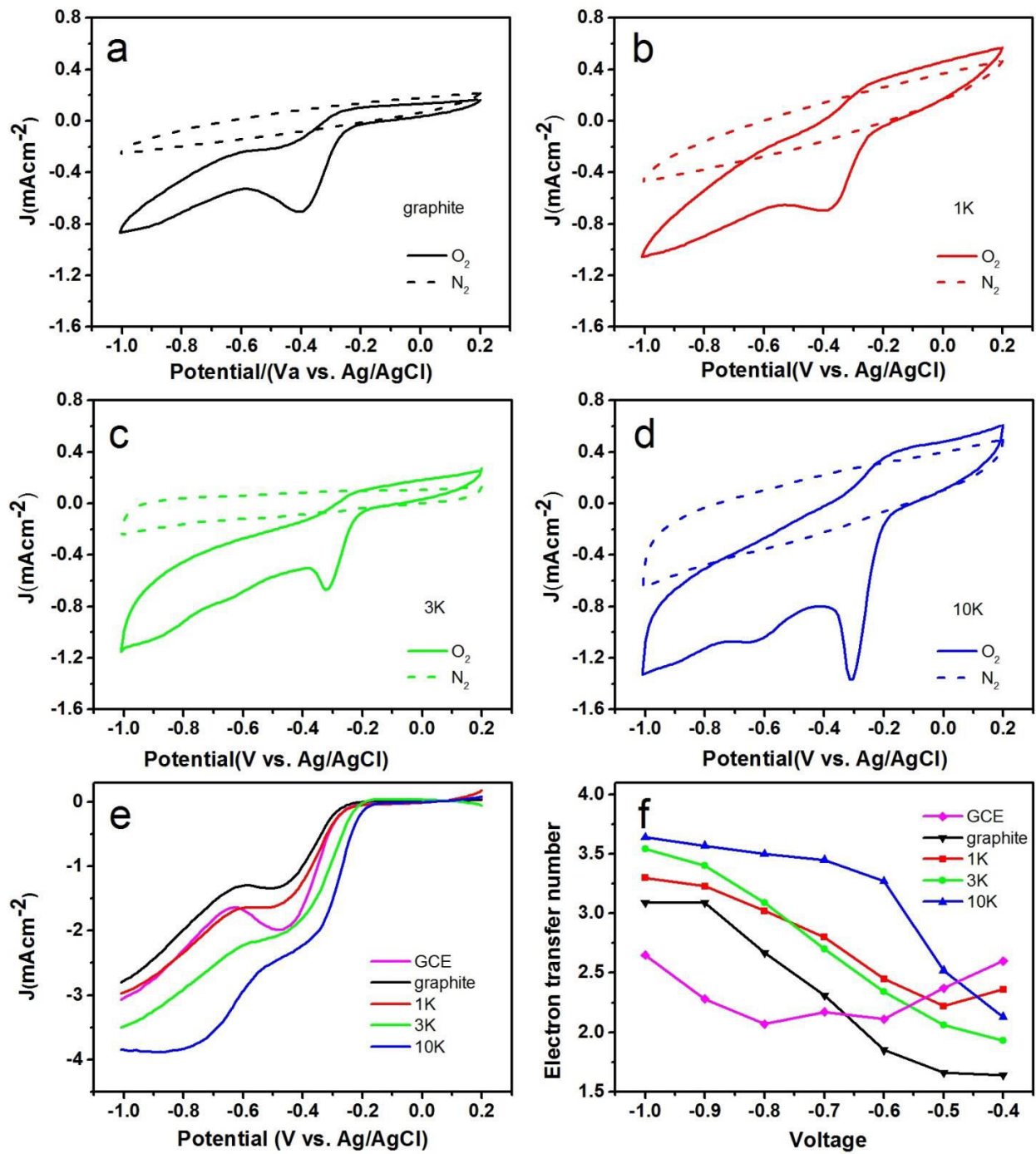

**Figure 4.** Cyclic voltammograms of catalysts in oxygen or nitrogen saturated 0.1M KOH at a scan rate of 100 mV s$^{-1}$. a) Pristine graphite. b) 1K. c) 3K. d) 10K. e) Linear sweep performance at a rotation speed of 1600rpm of the four catalysts in oxygen saturated 0.1M KOH at a scan rate of 10 mV s$^{-1}$. f) Electrons transferred per oxygen molecule for the four catalysts over potential

range -0.4V to -1.0V based on the Koutecky-Levich equation. For all electrochemical measurements a 3mm diameter BASI GCE was used and a catalyst loading of 0.283 mg/cm$^2$.

The ORR electrocatalytic activity of both starting graphite flakes and centrifugation products was firstly evaluated by cyclic voltammetry (CV) in $N_2$ and $O_2$ saturated 0.1M aqueous solution of KOH, as shown in Figure 4a-d. Comparative studies were performed on all catalysts with the same mass loading of 0.283 mg/cm$^2$. Conditioning of 10K catalyst is displayed in figure S12, showing the 1$^{st}$ and 20$^{th}$ CV scan. Voltammograms without any peaks were obtained in the absence of oxygen for all catalysts. A relatively increased quasi-rectangular shape was obtained for the 10K graphene nanosheets due to the increased surface area associated with their small lateral size and reduced number of layers. When $O_2$ was introduced, a typical cyclic voltammograms of graphite showed one broad reduction peak at –0.40 V, which is associated with a two-electron electrochemical reduction. In contrast, the 10K product showed two reduction peaks located at –0.31 and –0.66 V, which are associated with two successive two-electron electrochemical reductions. The reduction peak at –0.31 V is attributed to the reduction process of $O_2$ to $HO_2^-$, while the peak at –0.66 V corresponds to the reduction of $HO_2^-$ to $OH^-$. Similarly the 3K product showed two ORR peaks but of lower intensities centered at more negative potentials of -0.40 and -0.66 V, respectively, (Figure 4c). The main cathodic peak potential for the 10K product is more positive than that of various nitrogen doped graphene[53, 54] and close to the cathodic peak of edge-nitrogenated graphene nanoplatelets (-0.28 V)[55]. These results highlight the low overpotential for ORR on nanosized graphene with a highly crystalline basal plane, free from metal impurities and edge functionalized groups with only a low amount of oxygen content.

To gain further insight into the role of abundant edges on the ORR electrochemical process the electrocatalytic performance of the above 4 catalysts was examined by linear sweep voltammetry (LSV) in an aqueous solution of $O_2$-saturated 0.1M KOH, at a rotation rate of 1600 rpm and a scan rate of 10 mV/s on a 3mm diameter BASI GC electrode. The effect of mass loading for the 10K catalyst on LSVs is presented in Figure S13. LSVs presented in Figure 4e, have been corrected by subtracting the background response obtained in $N_2$ saturated 0.1M KOH solution. Compared to graphite, the 1K, 3K and 10K products are significantly more active in terms of lower overpotential and larger current density (Figure 4e and Figure S14a). The LSVs reveal an appreciable change on the onset potential between graphite (-0.19 V) and centrifugation products (1K: -0.19, 3K: -0.16, 10K: -0.13 V) as well as stronger diffusion current densities (10 K: -3.85 mA/cm$^2$) through the whole potential range as the centrifugation speed increases. A higher population of edge defects is present on the centrifugation products compared to that on graphite, and their surplus has a substantial effect on the onset potential. Remarkably, the 10K electrode exhibited a more positive onset potential and a much larger limiting current for the ORR in 0.1 M KOH solution, compared to the edge functionalized (with hydrogen, carboxylic acid, sulfonic acid and carboxylic acid/sulfonic acid) graphene nanoplatelets prepared through a ball milling technique[56] or reduced graphene oxide[57]. The performance is similar to macroporous graphitic carbon nitride/carbon composite (g-$C_3N_4$/C: onset: -0.14 V; at 1V J~4 mA/cm$^2$)[58], and comparable to that of three–dimensional nitrogen–doped graphene aerogel–supported $Fe_3O_4$ nanoparticles ($Fe_3O_4$/N–Gas: onset- -0.19 V; J=4.2 mA/cm$^2$)[55], nitrogen doped graphene infused with iron nanoparticles (NG/Fe$_5$: onset=-0.04 V; at 1 V J=4 mA/cm$^2$)[56].

These result highlight the improved oxygen reduction ability of 10K catalyst, which is clearly due to the following synergetic effects: (i) the abundance of edges capable to influence the adsorption and dissociation of $O_2$ and (ii) the efficient electron transfer between the electrode and active edge sites ensued from the reduced number of layers and the high structural crystalline quality of the basal plane. It is known that both graphite and a few layer graphene exhibit anisotropic electron conduction in directions parallel and perpendicular to the basal plane. The electron transfer in the vertical direction decreases as the number of layers increases. In thick graphite platelets, even though a rich amount of edge sites are exposed in the direction perpendicular to the basal plane their electrocatalytic performance is poor due to the increased electron path.

Since all the centrifugation products and even graphite have similar oxygen contents, it becomes obvious that the gradually enhanced performance is not caused by the limited oxygen-sites[59, 60], but is the outcome of a simultaneous reduction in the lateral size and progressive reduction in the number of layers. The onset potential of -0.13 V vs. Ag/AgCl and the current density of -3.85 mA/cm$^2$ at 1 V, represent the best ORR performance ever achieved from an undoped carbon based catalyst.

Figure S14b shows the RDE voltammograms for the ORR on the 10K electrode at various rotation speeds (from 400 to 2025 rpm). With an increase of the rotation speed, the reduction current increases, which can be explained by shortened diffusion distance at high speeds. The Koutecky– Levich plots ($j^{-1}$ vs. $\omega^{-1/2}$) at different electrode potentials displayed good linearity (Figure S15), and thus, the slope of these Koutecky–Levich plots could qualitatively give a good prediction on the catalytic activity for $O_2$ reduction. The highest slope was obtained at 10K catalyst electrode, again suggesting its higher catalytic activity for ORR. The electron-transfer numbers (n) of the four electrodes at different potentials were calculated according to the slopes of the linear fitted Koutecky–Levich (K-L) plots ($J^{-1}$ vs, $\omega^{-1/2}$) and are presented in Figure 4f.

Generally, increasing n values from 1.64 to 3.09 were obtained for graphite, 1K and 3K nanosheets as the potential became more negative. This observation, reveals that peroxide is formed first at low overpotentials, which is then partially reduced further to –OH at higher overpotentials, during the oxygen reduction process. In contrast the 10K product exhibited much higher n values through the whole the whole potential range compared to graphite, 1K and 3K products. Based on the n values the electrocatalytic activity was in the order: 10K>3K>1K>Graphite. Over the potential range -0.7 V to -1 V the n values for the 10K electrode is almost stable (3.6-3.4) indicating that the reduction takes place through an "*apparent*" four electron transfer reaction, due to faster hydrogen peroxide decomposition by the electron rich edges. Clearly, our results indicate that a substantial improvement in the electrocatalysis of ORR can be brought by a simultaneous reduction in both the lateral size and number of layers of graphene nanosheets. These changes bring an increase in the number of catalytic active sites and efficient transfer of electrons both of which synergistically enhance the ORR activity.

The reduction of molecular oxygen can proceed through two major pathways under alkaline conditions. A direct four electron pathway is the preferred route, producing the desired $OH^-$ described by equation (1). Alternatively ORR can go through a 2e⁻ +2e⁻ pathway involving the production of a peroxide $HO_2^-$ intermediate in the first 2e⁻ process (equation (2)), and subsequently undergo a further 2e⁻ reduction (equation (3))[61]. If reactions (2) and (3) are very fast, it would then appear as if oxygen is reduced directly to $OH^-$ through the four electron transfer process.

$$O_2 + 2H_2O + 4e^- \rightarrow 4OH^- \tag{1}$$

$$O_2 + H_2O + 2e^- \rightarrow HO_2^- + OH^- \tag{2}$$

$$HO_2^- + H_2O + 2e^- \rightarrow 3OH^- \tag{3}$$

For our electrochemical findings, it would appear that for the 10K catalyst, the intermediate product of the first 2e⁻ process of the ORR (i.e. first two-electron reduction to hydrogen peroxide at low overpotential) has as much better chance to undergo the second 2e⁻ process compared with the rest of the catalysts examined.

The hydrogen peroxide yield ($H_2O_2$ %) and the number of electrons transferred (n) per oxygen molecule in the ORR were also estimated using Rotating Ring Disk (RRDE) measurements (Figure S16). An agreement between the K–L and the RRDE methods for the 4 catalysts studied (graphite, 1K, 3K 10K) was only possible for the 10K over the range -6V to -1V. Due to the absence of a diffusion limiting region in the rest of the catalysts no agreement between the calculated n numbers by two methods was observed. It is generally accepted that the RRDE method, which is able to directly determine the amount of peroxide species produced, generates more accurate results[62].

In order to evaluate the contributions to the electrochemical activity due to increased density of active sites per surface area, we employ a Tafel analysis in conjunction with electrochemical double layer capacitance measurements. Tafel analysis of the j-v curves of the four samples (graphite, 1K, 3K, 10K) exhibited slopes over the range 84 to 77 mV per decade suggesting a similarity of surface chemistry for the ORR. Comparison of the relative activity of the four samples in terms of the exchange current densities, $j_0$ (determined from the extrapolated x-intercept where V = 0), revealed the following trend Graphite>1K>3K>10K, as seen in Figure S17.

The relative electrochemical surface active area of the catalysts was estimated by measuring the electrochemical double-layer capacitance ($C_{DL}$) over the potential range of -0.1 to 0.1 V vs. Ag/AgCl. This potential range contains no faradaic current and is associated with the double layer charging. The charging current, $i_c$, is then measured at various scan rates, $\nu$, between 0.1 to 0.01 $Vs^{-1}$, (Figure S18). The $C_{DL}$, as given by equation (4)[58, 63].

$$J_c = \nu \cdot C_{DL} \tag{4}$$

Thus a plot of $j_c$ versus $\nu$ gives a straight line with a slope equal to the $C_{DL}$ as shown in Figure S19. The estimated $C_{DL}$ values are given in table S5. The analysis reveals a 1.6-3.7x increase in the density of active sites per surface area for the various centrifugation products compared to graphite, enabling the 10K to outperform by almost 4 times.

Theoretical calculations have shown carbon atoms in graphene that possess high spin or charge density are the effective electrocatalytic active sites for ORR[19]. The basal plane of pure graphene does not possess any asymmetry in charge or spin density and therefore does not have catalytic activity for ORR. Recently functional density calculations have predicted that carbon atoms at the zigzag edge possess high positive charge density, which can act as catalytic active sites for ORR[64]. Therefore a plethora of zigzag edges could provide more active sites, where the carbon atoms with high charge density follow a four-electron transfer pathway. This hypothesis is supported by our findings, where abundant edges on the 10K nanosized graphene result in enhanced ORR activity. In general exfoliated graphene flakes appear to show zigzag and armchair edges with comparable probability[25, 65]. Therefore an increase in the overall edge sites would also increase the population of active zigzag edges.

Even though an abundance of zigzag edge sites is beneficial, the layer number should also be optimized. During ORR, the electron transfer has to proceed between the electrode support and the edges of all successive layers, which leads to significantly increased resistance. Resistivity has been measured to be larger vertical to the basal graphene planes as compared to parallel to the planes[66]. This means that the single layer or a few layer nanosized graphene is the optimal structure compared to a multi-layer nanosized graphene, because of the shorter electron path.

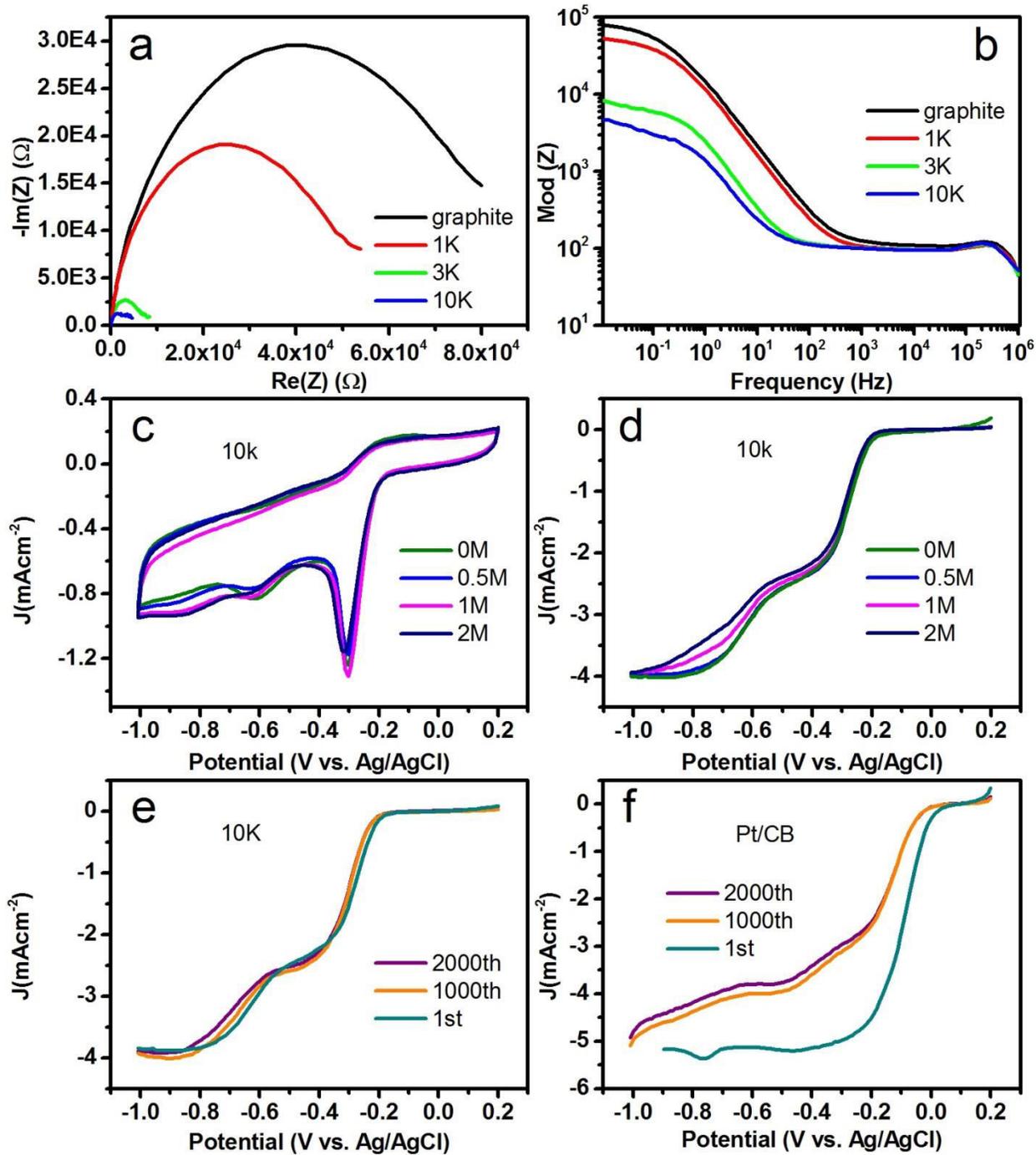

**Figure 5.** Electrochemical impedance spectroscopy plots in the form of a) Nyquist and b) Bode plots of the four catalysts Measurements were performed at -0.2 V. c) CV, d) LS response of 10K nanosheets after methanol addition. e) LS curves of 10K, f) benchmark Pt/CB (20 wt % Pt)

after 1st, 1000th and 2000th CV scans. All measurements were carried out in oxygen saturated 0.1M KOH using a 3mm diameter BASI GCE and a catalyst loading of 0.283 mg/cm$^2$.

This sequence was further confirmed by the electrochemical impedance spectra (EIS), which can be used to probe the interfacial processes and kinetics of electrode reactions in electrochemical systems. EIS measurements were performed at a potential -0.2 V for the 4 catalysts. The Nyquist plots, shown in Figure 5a, which present impedance values in the form of imaginary ($Z_{Im}$) vs. real ($Z_{Re}$) parts at various frequencies, showed an arc-like profile for all four catalysts. As graphite possesses a low ORR activity, this is mirrored with a large charge transfer resistance as it does not allow fast shuttling of electrons during ORR. The 10K exhibits the smallest semicircle, which clearly demonstrates that the 10K electrode possesses significantly lower charge transfer resistance and thus allows a much faster transport of electrons during ORR. The overall magnitude of the impedance in the low frequency range as determined by the intercept of the straight lines with the vertical axis (modulus of Z) in the Bode plot shown in Figure 5b, follows the order Graphite>1K>3K>10K. We rationalize the larger decrease in impedance for 10K in terms of its earlier onset potential and hence higher catalytic activity.

Tolerance to crossover has been investigated through the introduction of methanol to the O$_2$ saturated KOH electrolyte. Both graphite and 10K product demonstrate very little alteration in both CV responses and LS voltammograms compared to commercial platinum carbon black (Pt/CB) upon introduction of various concentrations of methanol as shown in Figure 5c-d and supporting Figure S20,. This demonstrates their high catalytic selectivity for the ORR against methanol oxidation. The ORR activities of starting graphite, 10K and Pt/CB (0.283 mg/cm$^2$) before and after 2000 cycles are shown in Figure S21 and Figure 5e-f, respectively. There is only 15 mV and 24 mV shift of the first half-wave potential ($E_{1/2}$) for graphite and 10K respectively.

By comparison, the shift of $E_{1/2}$ is 107 mV for commercial Pt/CB after 2000 cycles, revealing the superior durability of the 10K product.

Through this study by investigating progressively thinner and smaller graphene nanosheets with minimal oxygen content and no metal contamination, new insights are gained into the roles of graphene edges and thickness on the ORR activity. Experimentally, it was shown that the lateral size reduction of few-layer graphene effectively elevates the electrode onset potential and current density, resulting in enhancement of the ORR activity. There is currently an increasing notion that the activity and reaction kinetics for the ORR are strongly correlated with the work function (WF) of the electrocatalyst[67, 68]. A decrease in WF is associated with an enhancement in ORR activity. Theoretical studies predict and experimental studies have verified that that the WF of graphene is in a similar range to that of graphite, ~4.6 eV[69], however it depends sensitively with the number of layers and lateral size[69, 70]. It has been predicted that the WF decreases with a reduction in the number of layers and lateral size. Therefore the progressive enhancement in ORR for the graphene products can be explained through a progressive decrease in the WF.

**Conclusion**

We have developed a simple and versatile ionic liquid assisted process coupled with sequential centrifugation steps to efficiently exfoliate the pristine graphite directly into a few layer nanosized graphene with low oxygen content (< 2.5±0.2 at %). We have examined for the first time their catalytic activity for oxygen reduction reaction. We found that the nanosized graphene exhibits an enhanced ORR performance, which is attributed to the synergy of two main factors (i) the abundance of edges sites accrued from the small lateral size and (ii) the efficient electron transfer between the active edge sites and the electrode. The last was ensued from the reduced number of graphene layers and the high structural crystalline quality of their basal plane. The progressive increase in ORR activity is a clear indication that the improvement is merely due to the size and thickness reduction and not an artifact from metal contamination or functional groups. The approach presented herein is applicable to the synthesis of other 2D nanomaterials. The size and thickness dependent catalytic activity of graphene nanosheets has important implications in understanding and further improving the catalytic activities of metal free graphene and possibly other 2D nanomaterials. While the ORR performance is not comparable to commercial Pt/CB catalysts, this work could offer opportunities for stimulating future experimental and theoretical research on the role of sheet size and number of layers of pristine graphene on electrocatalysis.

AUTHOR INFORMATION

*To whom correspondence should be addressed, p.papakonstantinou@ulster.ac.uk

The authors declare no competing financial interest.

ACKNOWLEDGMENT

This work was financially supported by a Leverhulme Trust/Royal Academy of Engineering Senior Research Fellowship to P.P; A PhD Studentship to JB from the Department of Education in Northern Ireland; A visiting Senior Research Fellowship to M.L. by the University of Ulster; The Chinese Thousand-Talents plan program and the Jiangsu Shuangchuang program (Q. X and P. W.).

**TOC Figure.**

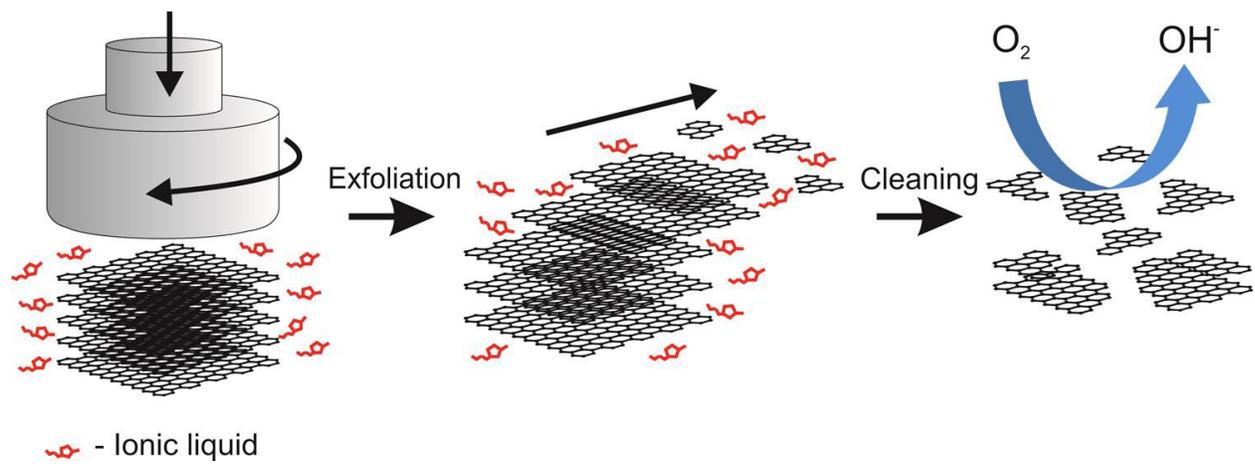

# Tuning the catalytic activity of graphene nanosheets for oxygen reduction reaction via size and thickness reduction

*John Benson[a], Qian Xu[b], Peng Wang[b], Yuting Shen[c], Litao Sun[c], Tanyuan Wang[d], Meixian Li[d], and Pagona Papakonstantinou\*[a]*

[a] School of Engineering, Engineering Research Institute, University of Ulster, Newtownabbey BT37 0QB, UK

[b] National Laboratory of Solid State Microstructures and College of Engineering and Applied Sciences, Nanjing University, 22 Hankou Road, Gulou, Nanjing, 210093, P. R. China

[c] SEU-FEI Nano-Pico Center, Key Laboratory of MEMS of Ministry of Education, Southeast University, Sipailou 2, Nanjing 210096, P. R. China

[d] College of Chemistry and Molecular Engineering, Peking University, Beijing 100871, P.R.China.

*To whom correspondence should be addressed: p.papakonstantinou@ulster.ac.uk

## I. Synthesis of graphene products

Graphite (99.99% purity) was ground in a small amount of ionic liquid and cleaned in a dispersion of acetone and DMF. The cleaned ground graphite DMF dispersion was centrifuged at 1,000 rpm to collect aggregates. The DMF suspension was further centrifuged at 3,000 rpm to yield nanosheets with a higher degree of exfoliation compared to Precipitate 1K. The supernatant remaining after this centrifugation step was further centrifuged at 10,000 rpm showing the maximum degree of exfoliation. The precipitates by centrifugation at 1,000, 3,000 and 10,000 rpm were collected and subjected to analysis in order to compare their degree of exfoliation. The labeling scheme of the different nanosheet fractions is 1K, 3K and 10K respectively.

## II. Electrochemical Measurements

Inks of the respective catalysts were prepared by dispersing 5mg of catalyst material in 1 ml DMF (5mg ml$^{-1}$) and ultrasonicated for 1hr. 50µl of Nafion (Sigma Aldrich, Nafion 117 solution, Lot#70160) solution was then added to the catalyst ink and further sonicated for 10 minutes. A total of 4µl of well dispersed catalyst ink was drop dried onto a pre polished glassy carbon electrode (3 mm diameter, BASI) giving a loading of 0.283 mg/cm$^2$. The ink was dried under an IR lamp. Testing of the prepared working electrodes was carried out using a software-controlled potentiostat (Autolab, PGSTAT20) with a typical three electrode electrochemical set up. A platinum wire was used as a counter-electrode, an Ag/AgCl (3M KCl) electrode as the reference electrode and glassy carbon electrode (3 mm) as a working electrode. Oxygen or nitrogen saturated 0.1M KOH was used as the electrolyte and all tests were performed under ambient conditions under potentials between -1.2 and +0.2 V at a 10 mVs$^{-1}$ scan rate (100 mVs$^{-1}$ for cyclic voltammetry measurements). Rotating disk electrode measurements were performed at rotating speeds from 400 to 2025 rpm using an RDE.

Kinetic analysis was carried out using the koutecky-levich equation (S1).

$$\frac{1}{j} = \frac{1}{j_k} + \frac{1}{j_{diff}} = \frac{1}{j_k} + \frac{1}{B\omega^{0.5}} \qquad (S1)$$

Where $j_k$ is kinetic current, $j_{diff}$ is the diffusion limited current density, $\omega$ is the electrode rotation in rpm and $B$ is the Levich slope, given by eqn. (S2).

$$B = 0.62nFC_O D_O^{2/3} v^{-1/6} \qquad (S2)$$

Here, $n$ is the number of electrons transferred per oxygen molecule, $F$ is faradays constant, $D_{O_2}$ is the diffusion coefficient of oxygen (1.93x10$^{-5}$ cm$^2$ s$^{-1}$), $C_{O_2}$ is concentration of oxygen in solution (1.26x10$^{-2}$ cm$^2$/s) and $v$ is the kinematic viscosity of the KOH electrolyte (1.009x10$^{-2}$ cm$^2$ s$^{-1}$).

Electrochemical impedance spectra (EIS) were measured using frequency response analyser software. EIS were measured under static O$_2$ saturated KOH $_{(aq)}$ solution at -0.2 V vs Ag/AgCl. The sinusoidal wave was measured with amplitude of 10 mV, and the frequency range was from 1 MHz to 10 mHz.

The electrochemical measurements with the rotating ring disk electrode (RRDE) were conducted on a Pine Wavedriver 20 Bipotentiostat (Pine research instruments, Durham, USA). A catalyst loading of 0.229 mg/cm$^2$ was deposited onto the bare Pt ring GCE (Pine inst., E6 series Pt ring, AFE6R1PT, 5 mm diameter). A standard three electrode electrochemical setup was used, using a platinum wire as a counter electrode, an Ag/AgCl (3M KCL) electrode was used as reference. Oxygen saturated 0.1M KOH electrolyte was used. Measurements were performed at a scan rate of 10 mVs$^{-1}$ at a rotation speed of 1600 rpm between the potentials -1.2V and +0.2 V. The ring potential was held at +0.5 V vs. Ag/AgCl. The hydrogen peroxide yield (H$_2$O$_2$) and the number of electrons (n) transferred per oxygen molecule were calculated using equations (S3) and (S4)[1], respectively.

$$n = \frac{4I_d}{I_d + (I_r/N)} \quad (S3)$$

$$H_2O_2\ (\%) = 100\ \frac{2I_r/N}{I_d + I_r/N} \quad (S4)$$

Where $I_d$ and $I_r$ denote the disk and ring currents, respectfully. The collection efficiency (N) was 0.256 for the AFE6R1PT. RRDE (Figure S14) and mass loading measurements (Figure S20) were performed on the AFE6R1PT 5mm electrode.

### III. Characterisation Methods

The low resolution TEM images of as-prepared catalysts were taken on a JEOL 2100. The high resolution images were taken on a FEI Titan 80-300 Cs-corrected TEM at an accelerating voltage of 80 KV to minimize radiation damage. TEM specimens were prepared by dropping 2 µl of well dispersed catalyst in DMF onto carbon micro-grids (Agar scientific, S147-3, holey carbon film 300 mesh Cu). The grids were then dipped into acetone to remove excess DMF and dried under ambient conditions. High-resolution XPS spectra were taken using a Kratos AXIS ultra DLD with an Al Kα (hv=1486.6 eV) x-ray source. Spectra were fitted to a Shirley background. Thermogravimetric analysis was conducted on a SDT Q600 (TA instruments, V20.9, build 20) in a nitrogen atmosphere with a heating rate of 5°C/min (24°C-1000°C) for all four catalysts. Raman spectra were obtained using a He-Ne (632.8nm) laser with a Labram 300 system. XRD was conducted on powdered samples with a Bruker AXS D8 discover with Cu-kα radiation (40kV, 20mA, λ = 1.5418Å). SEM images were taken on a Quanta 200 3D system at a 10.7 mm working distance. . Inductively coupled plasma mass spectrometry (ICP-MS) was performed on a MC-ICP-MS (VG AXIOM, UK).

For ICP-MS analysis, the samples (solid powder, about 10 mg) were put into Pt crucibles, and then were heated at a temperature of 900 °C in a muffle furnace for half an hour, cooled down to room temperature. 1 mL of concentrated nitric acid was added into the crucible to dissolve the sample, and then was transferred into 50 mL of volumetric flask to constant volume with high purity water.

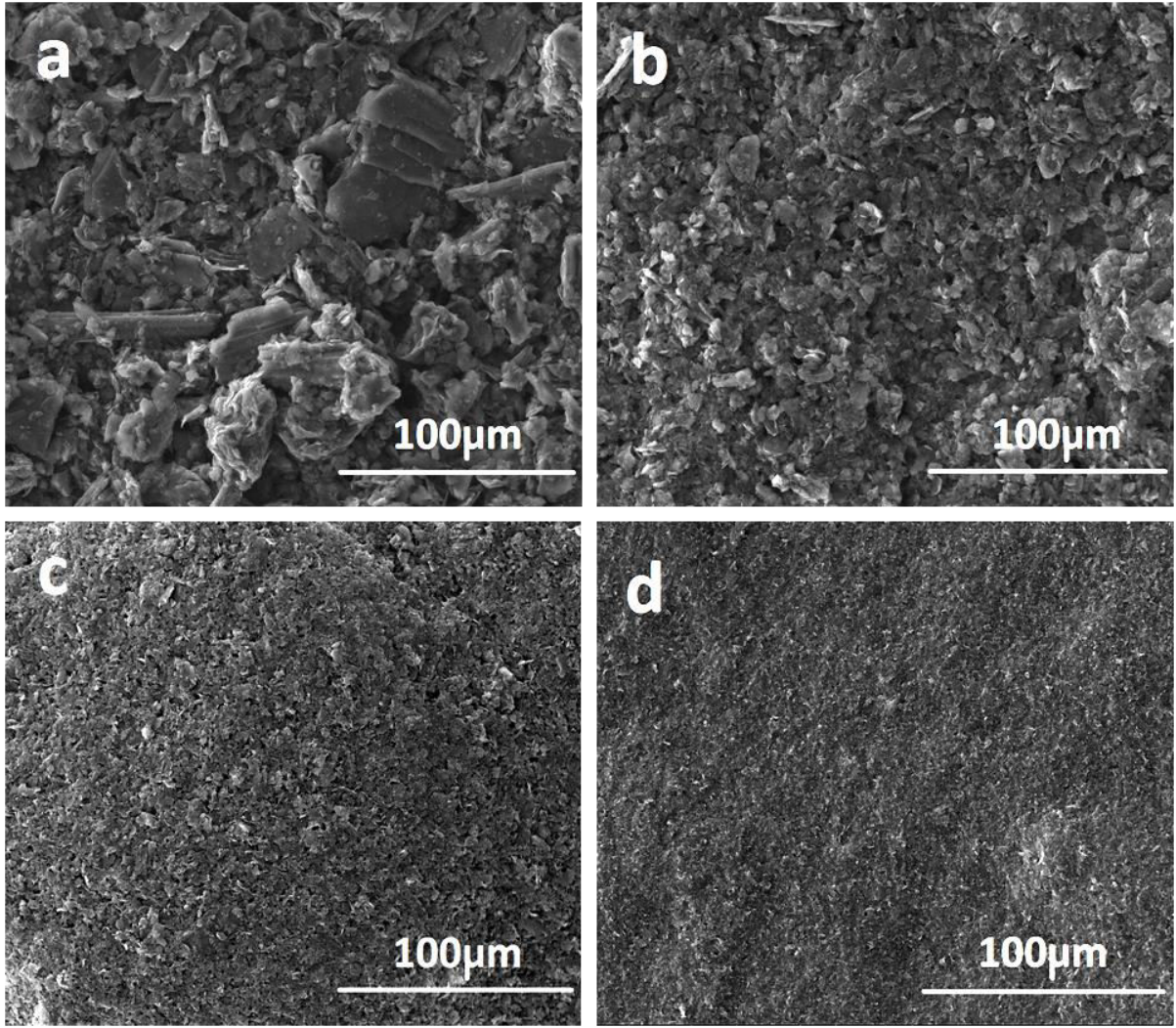

**Figure S1.** SEM images of a) graphite, b) 1K, c) 3K and d) 10K centrifugation products.

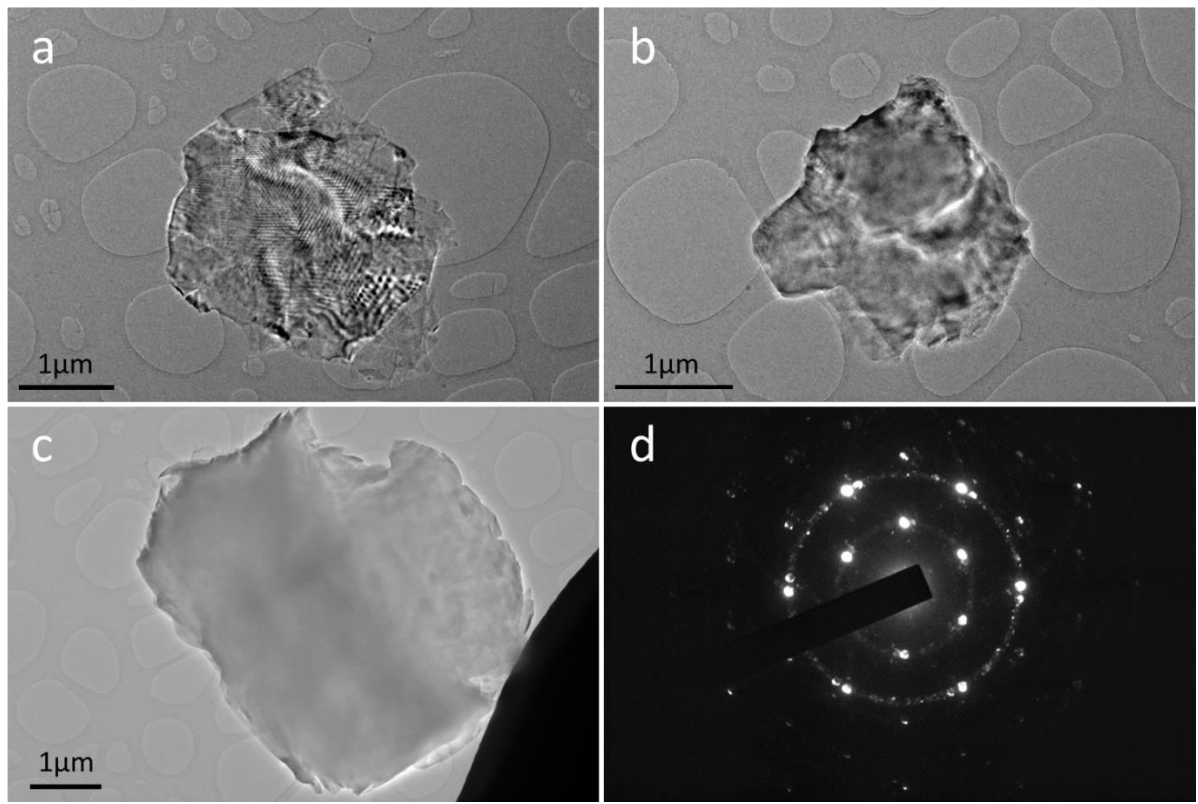

**Figure S2.** Low magnification TEM images from 1K sediment: a), b), c) panels are typical TEM images of objects found in the sediment produced by 1K. d) Is selected area electron diffraction (SAED) of the center of the region shown in panel a). The SAED pattern contains multiple sets of rotated spots in a hexagonal arrangement and of variable intensity, which could be caused by overlapping domains of graphene layers, back-folding of edges or intrinsic rotational stacking faults.

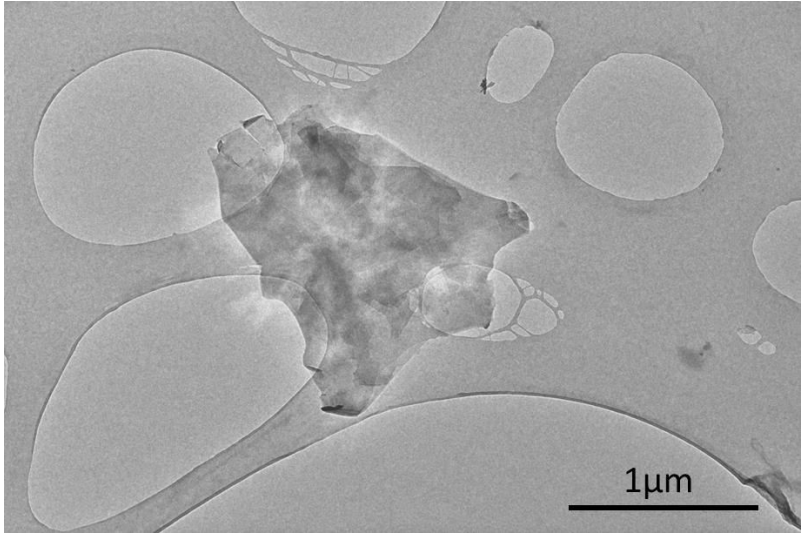

**Figure S3.** Low resolution TEM images from 3K sediment. Typical image of objects found in the sediment produced by 3K.

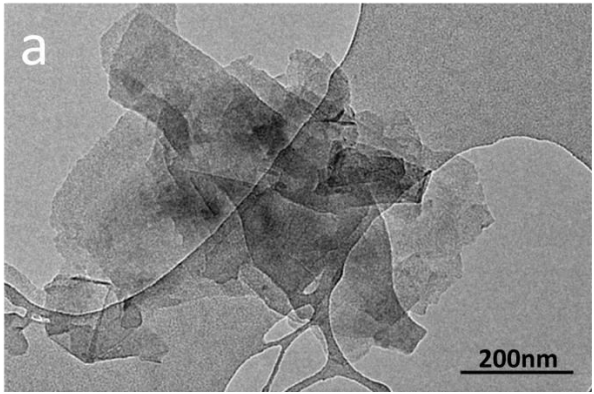

**Figure S4.** Low resolution TEM image from 10K sediment. Typical image of objects found in the sediment produced by 10K.

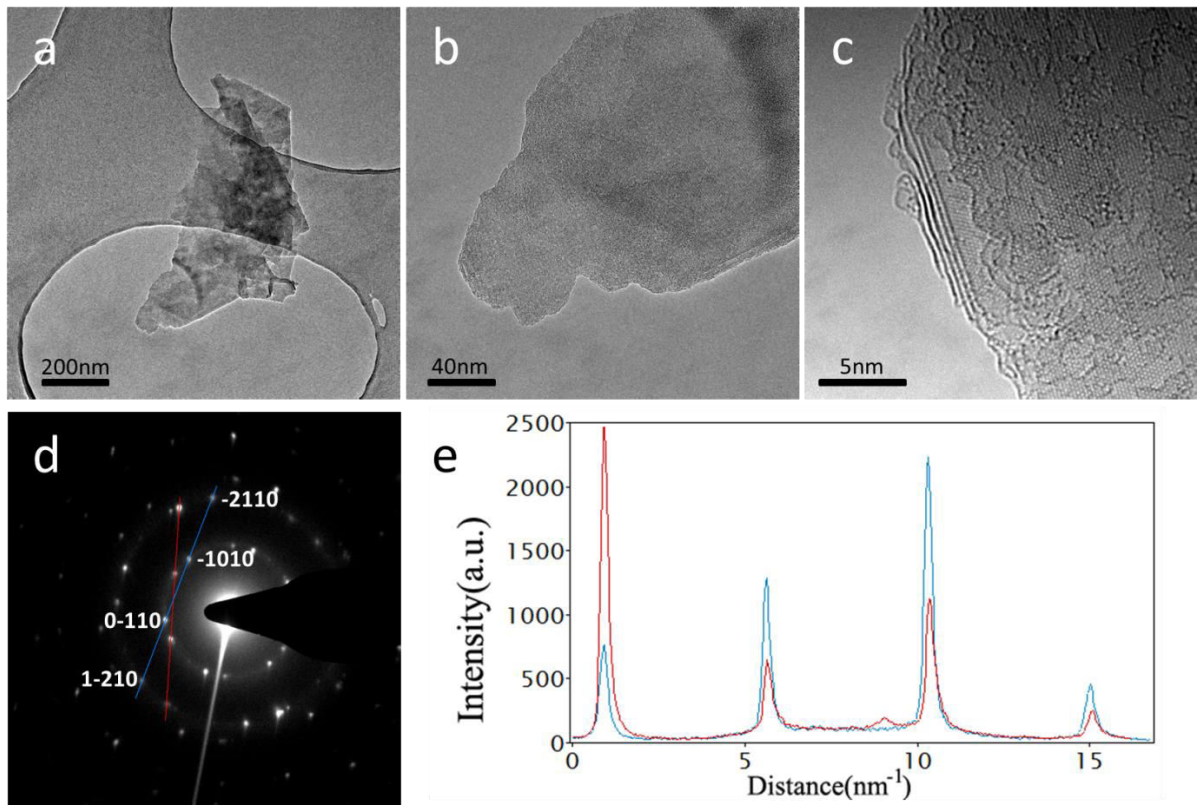

**Figure S5.** Low and high resolution TEM images from 10K: a), b), c) of a flake of about 5 layers thick. d) Is the corresponding electron diffraction pattern from c), indicating two different orientations of graphene. e) Diffracted intensity taken along the -2110 to 1-210 axis for the two oriented patterns in d). Ratio of the intensity of {1100} and {2110} is smaller than 1 in one orientation indicating multiple layers. While the ratio is larger than 1 in another orientation marked in blue indicating a single layer. Separation of fringes was measured to be 0.356nm.

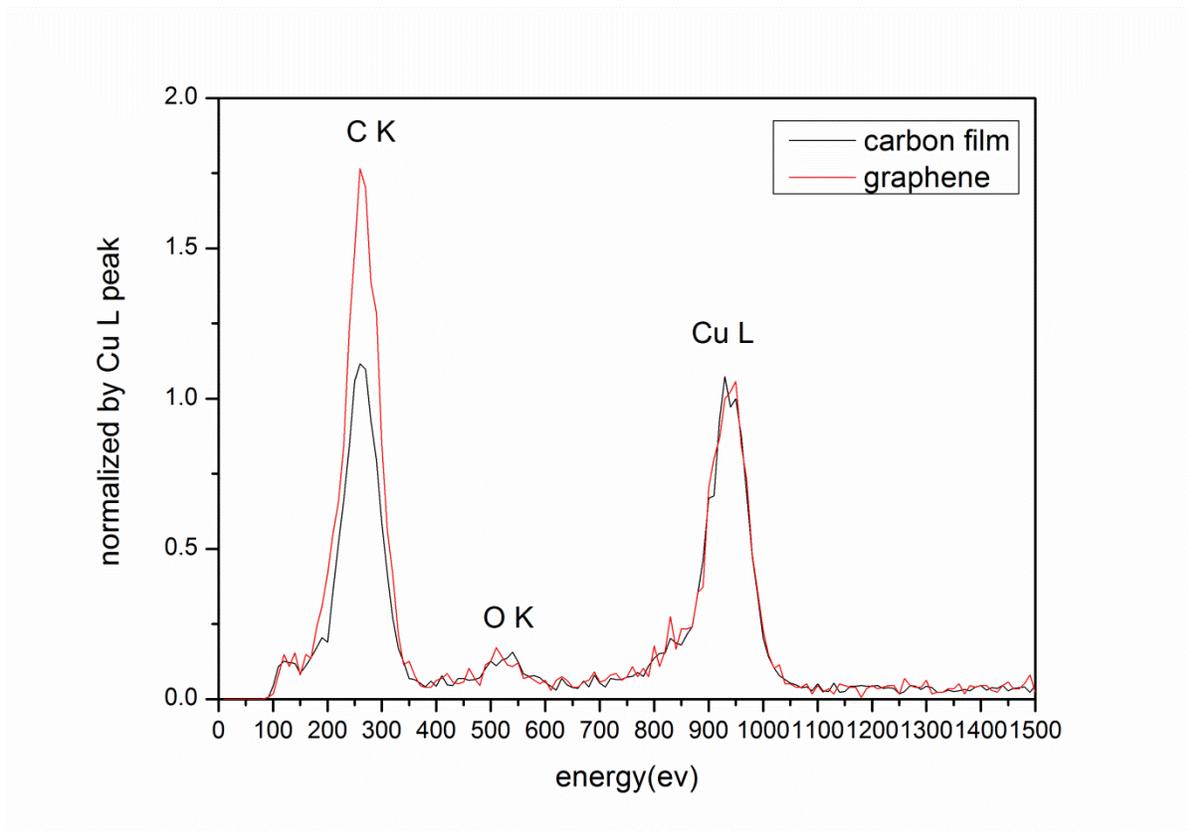

**Figure S6.** EDAX of 10K product compared with that of a carbon film. The small oxygen peak confirms the presence of only a small amount of oxygen in graphene. The absence of other foreign peaks indicates the high purity of graphene. Cu signal comes from copper grid.

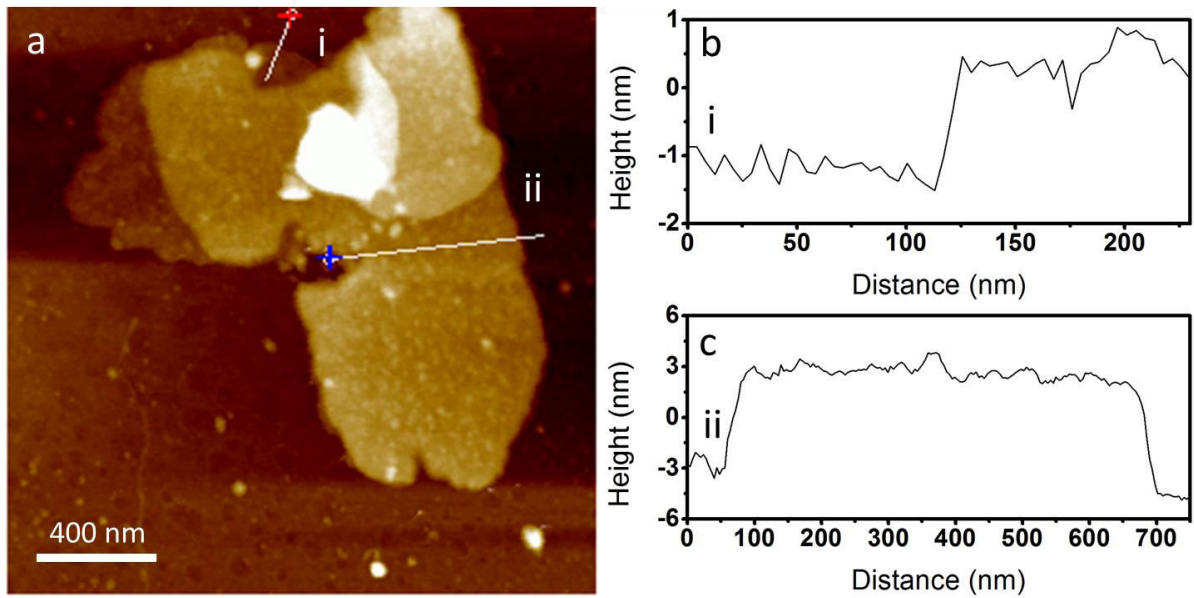

**Figure S7.** a) Representative AFM image of 3K product deposited on a silicon wafer, b) and d) subsequent height profiles for (i) and (ii), respectfully.

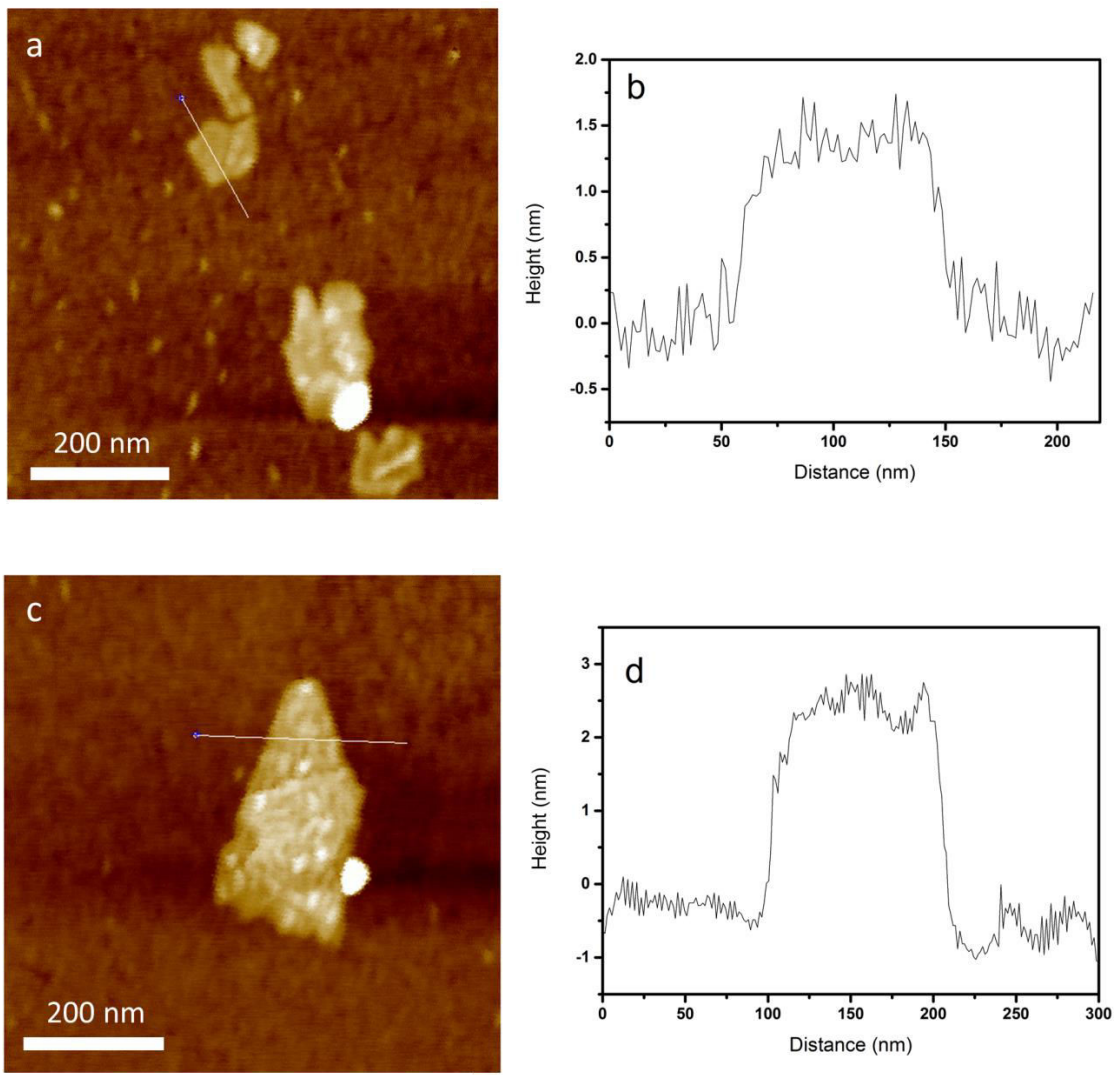

**Figure S8.** a) and c) representative AFM images of 10K catalyst deposited on a silicon wafer, b) and d) subsequent height profiles.

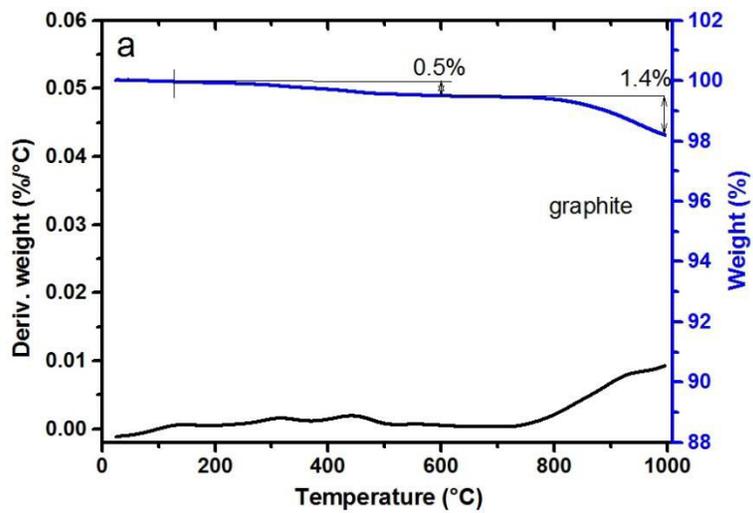
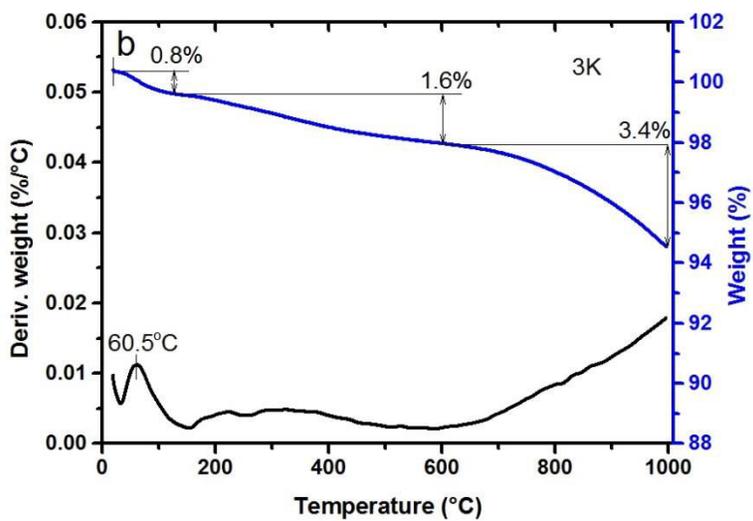
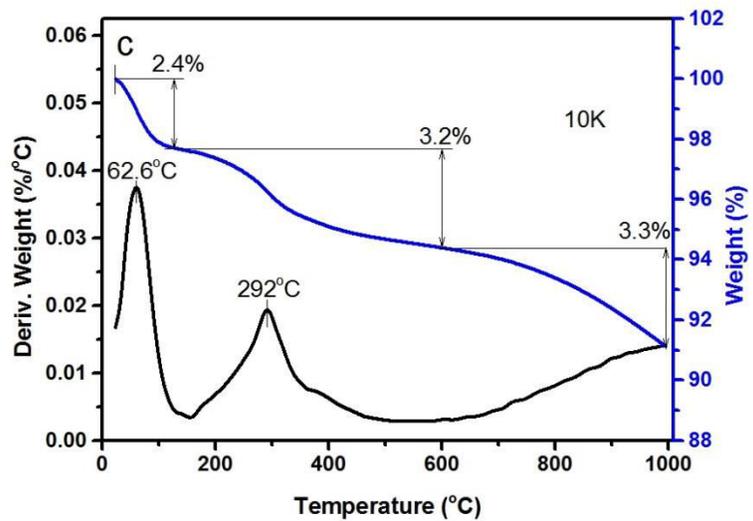

**Figure S9.** DSC-TGA plots of a) graphite, b) 3K and c) 10K. Data obtained at a heating rate of 5°C/min in nitrogen.

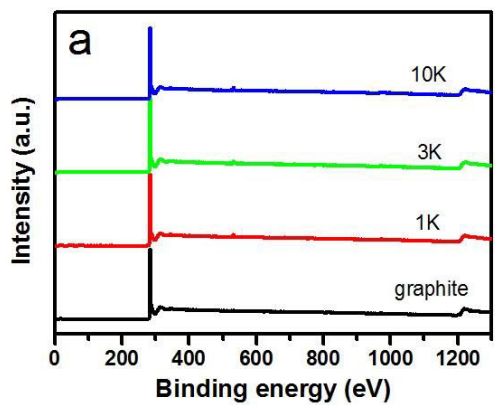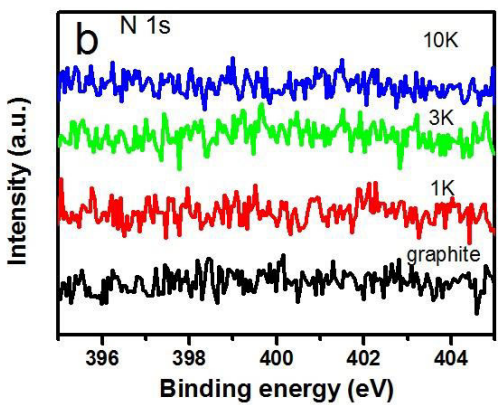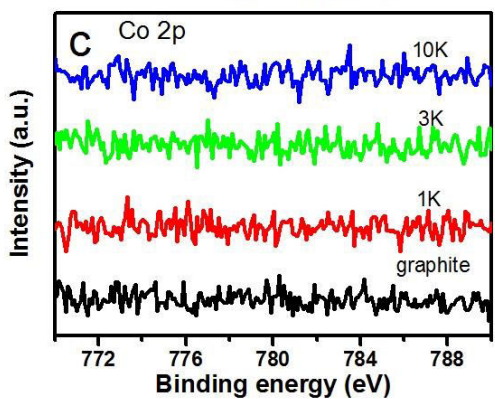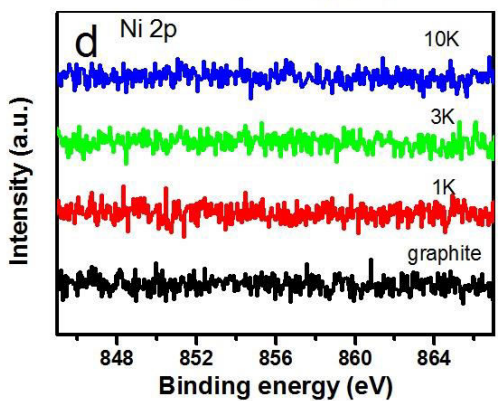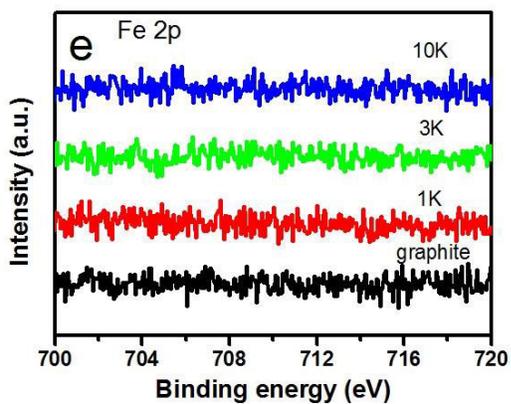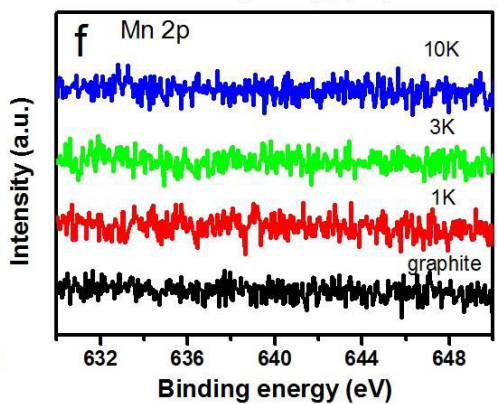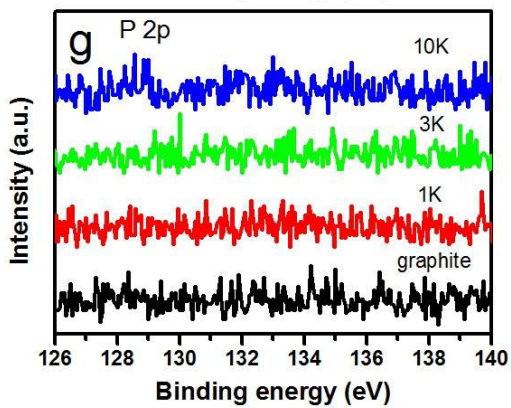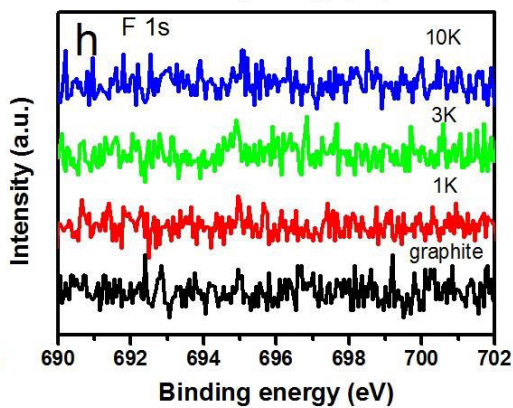

**Figure S10.** a) XPS wide survey spectra of graphite, 1K, 3K, and 10K centrifugation products showing no trace of contamination form used chemicals during exfoliation process. b) N1s, c) Co2p, d) Ni2p, e) Fe2p, f) Mn2p, g) P2p, h) F1s high resolution spectra reveal no contamination from used solvents or ionic liquid, indicating catalyst samples are of very high quality.

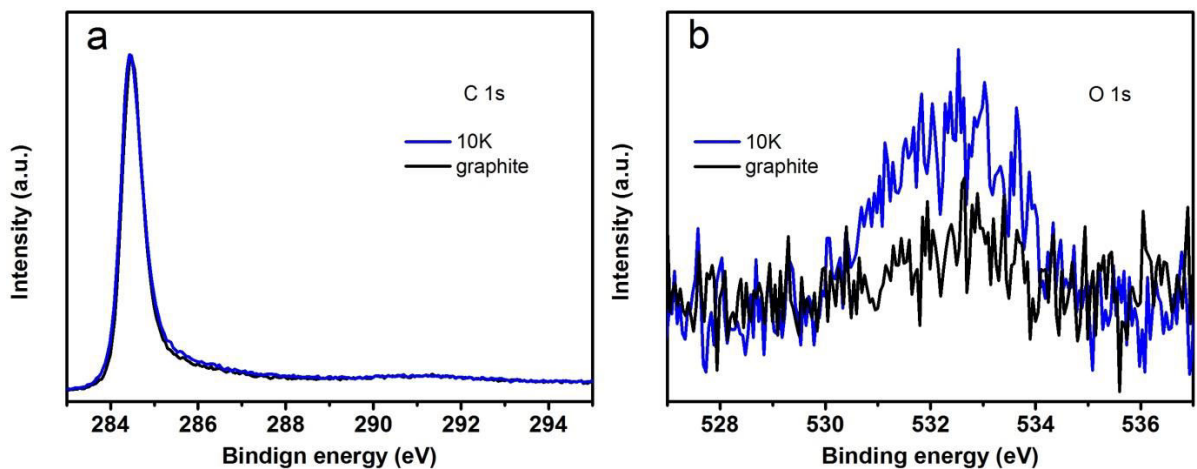

**Figure S11**. a) C 1s and b) O 1s overlap showing oxygen content.

High resolution C 1s and O 1s spectra reveal almost identical overlapping.

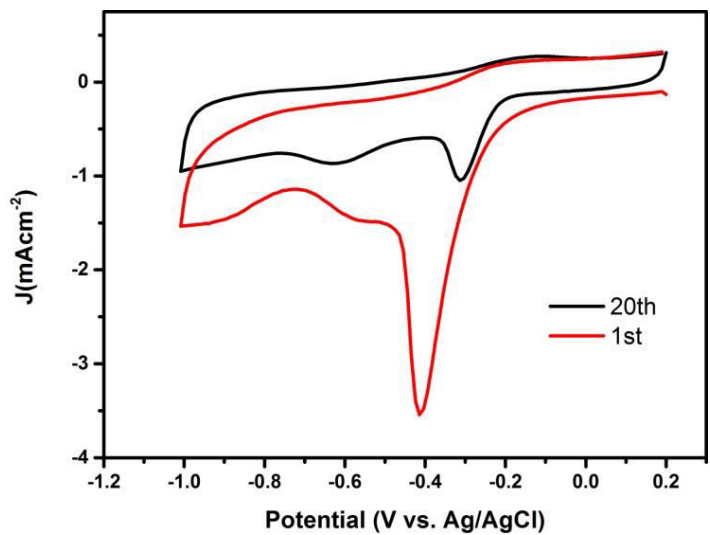

**Figure S12**. Cyclic voltammetry first and twentieth scan. Measurements performed in $O_2$ saturated KOH at a scan rate of 100 mV/s.

It is important to note that conditioning procedures can modify the initial state of the graphene edges. In our experiments 20 cyclic voltammetry (CV) scans were applied to the working electrode between -1 V and +0.2 V at 100 mVs$^{-1}$ to reach a steady CV curve. Figure S21 depicts the 1$^{st}$ and 20$^{th}$ scan. This conditioning of the surface is believed to decorate the active carbon edges with oxygen functional groups such as oxygen reduction intermediate OH $_{(ads)}$. Future synchrotron based high resolution XPS spectra will be pursued to confirm this assumption.

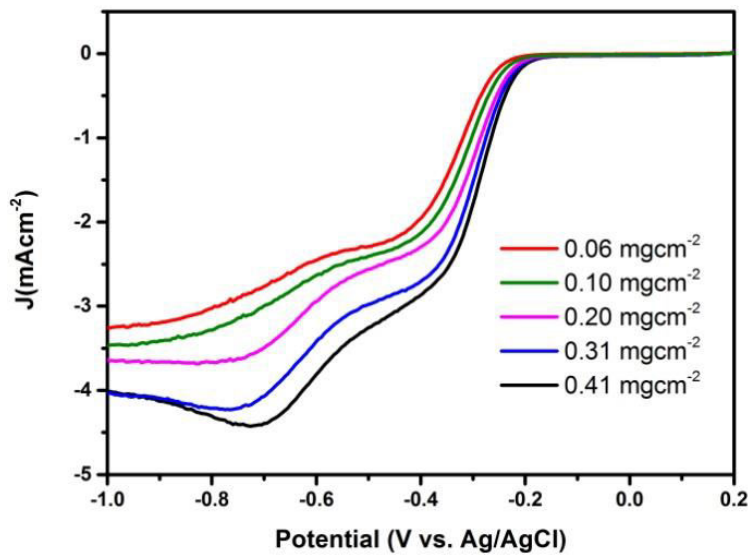

**Figure S13.** Linear sweep voltammetry curves showing effect of various mass loadings on 10K catalyst. Performed in $O_2$ saturated KOH electrolyte at a scan rate of 10 mV/s at 1600rpm (Pine. AFE6R1PT, 5mm electrode).

We have examined the effect of mass loading on the ORR performance of the 10K product. Figure S20 compares LS voltammetry curves for 10 to 80 µg mass loading which correspond to mass densities from 0.06 mg/cm$^2$ to 0.41 mg/cm$^2$. The ORR performance shows a slight increase in terms of onset potential from -0.16 V to -0.13 V. There is however an obvious increase in limiting current density as the mass of 10K catalyst increases. This behavior is in agreement with the well-known effects of mass loading on current density and onset potential for carbon materials[2].

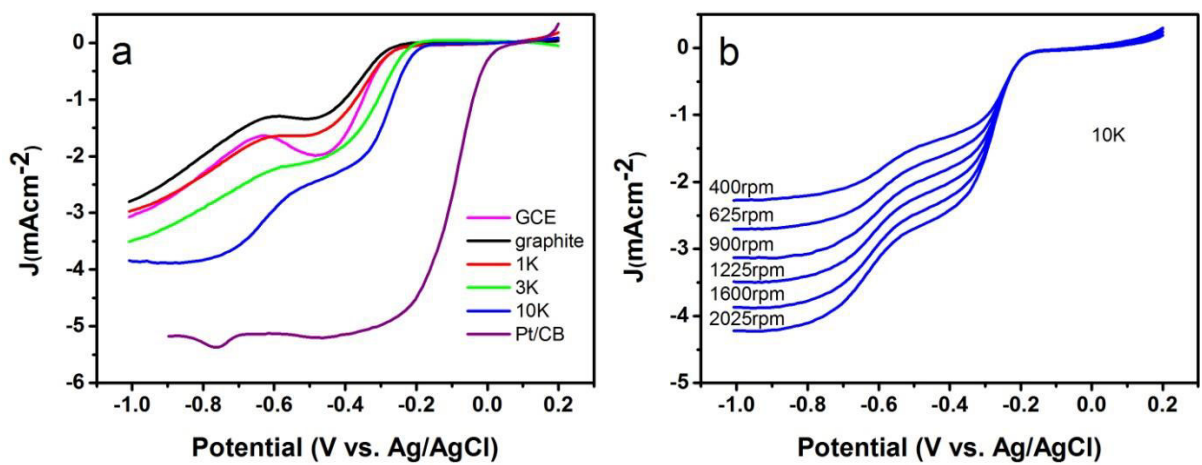

**Figure S14.** a) Comparison of four catalysts with commercial Pt/CB (20 wt % Pt) electrode (loading = 0.283 mg cm$^{-2}$) in $O_2$ saturated 0.1M aq. KOH at a rotation speed of 1600rpm (10mV s$^{-1}$ scan rate). b) Rotating disk electrode voltammograms at various rotation rates for 10K catalyst under $O_2$ saturated 0.1M aq. KOH (10mV s$^{-1}$ scan rate, BASI, 3mm diameter)

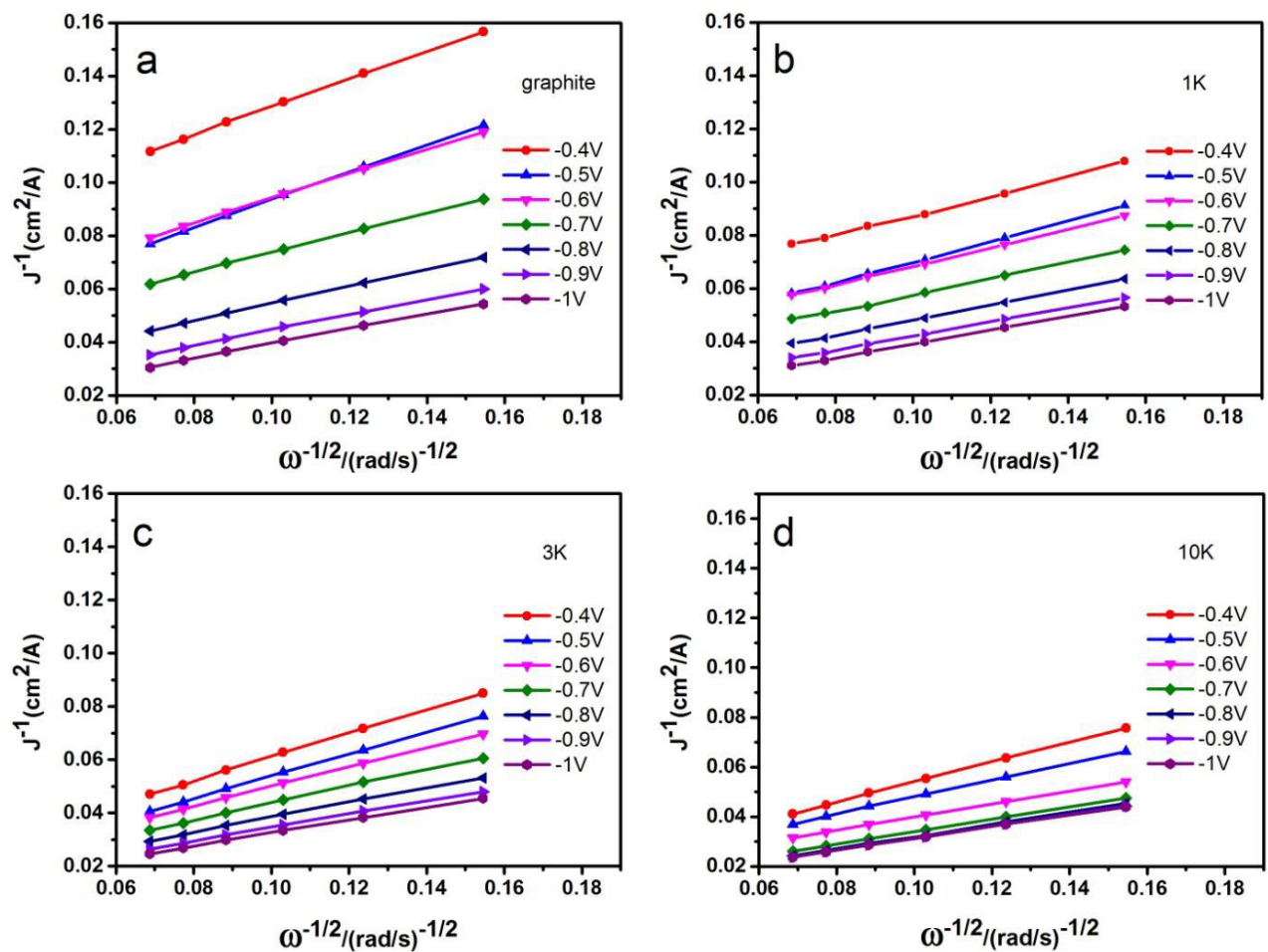

**Figure S15.** Koutecky-Levich plots of $j^{-1}$ *vs.* $\omega^{-1/2}$ of a) Graphite, b) 1K, c) 3K, d) 10K. Plots obtained from RDE data between voltages -0.4V to -1V.

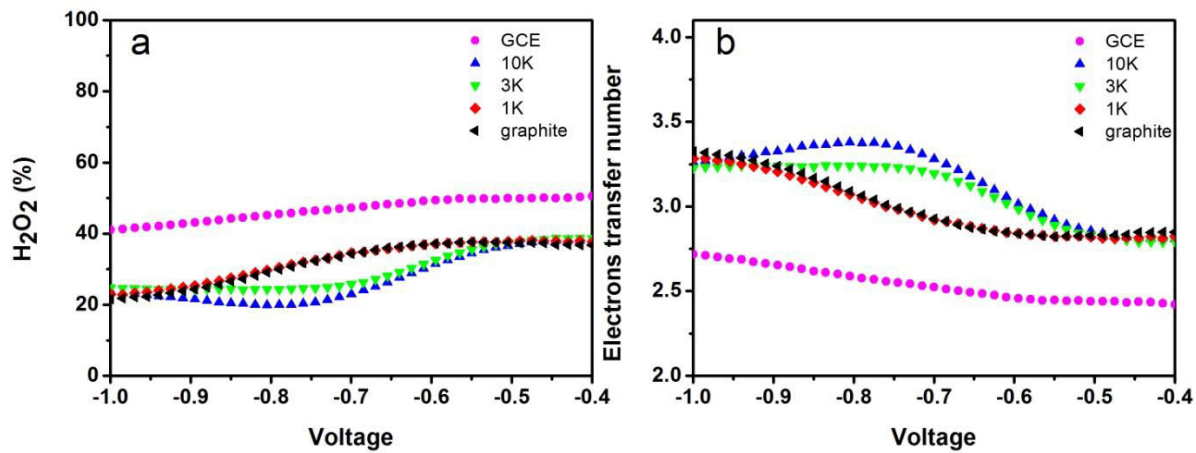

**Figure S16**. (a) Electron transfer numbers, (b) peroxide percentage produced at different potentials derived from the RRDE data from all catalysts and bare Pt ring GCE. Measurements performed in $O_2$ saturated KOH at a scan rate of 10 mVs$^{-1}$ (5mm diameter Pt ring GCE).

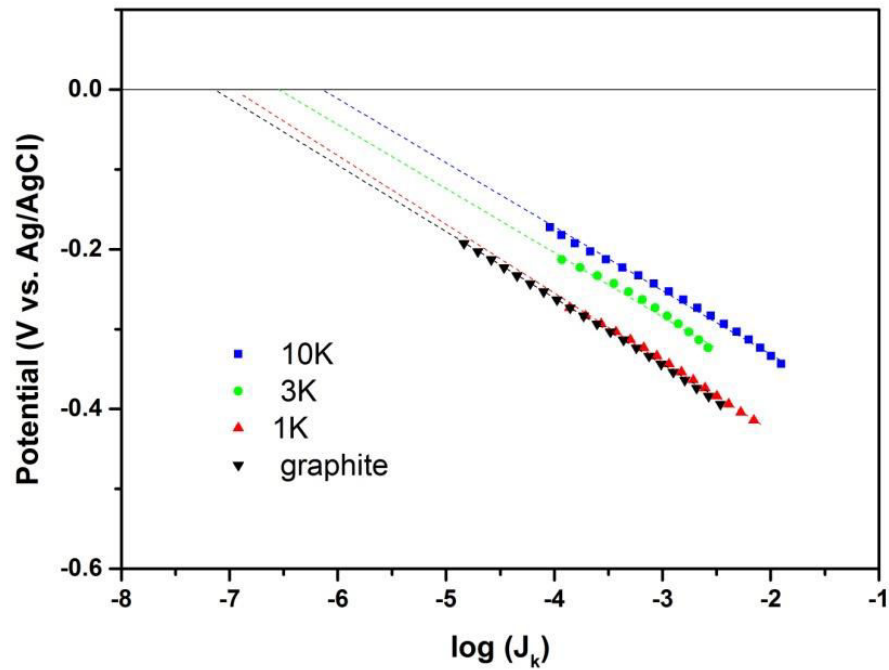

**Figure S17.** Tafel plots of four catalysts extrapolated to determine the exchange current densities. $J_k$ corresponds to the kinetic current density and was determined using the following equation (S5).

$$J_k = (J_{df} * J)/(J_{df} - J) \tag{S5}$$

Where $J_{df}$ is the diffusion current density.

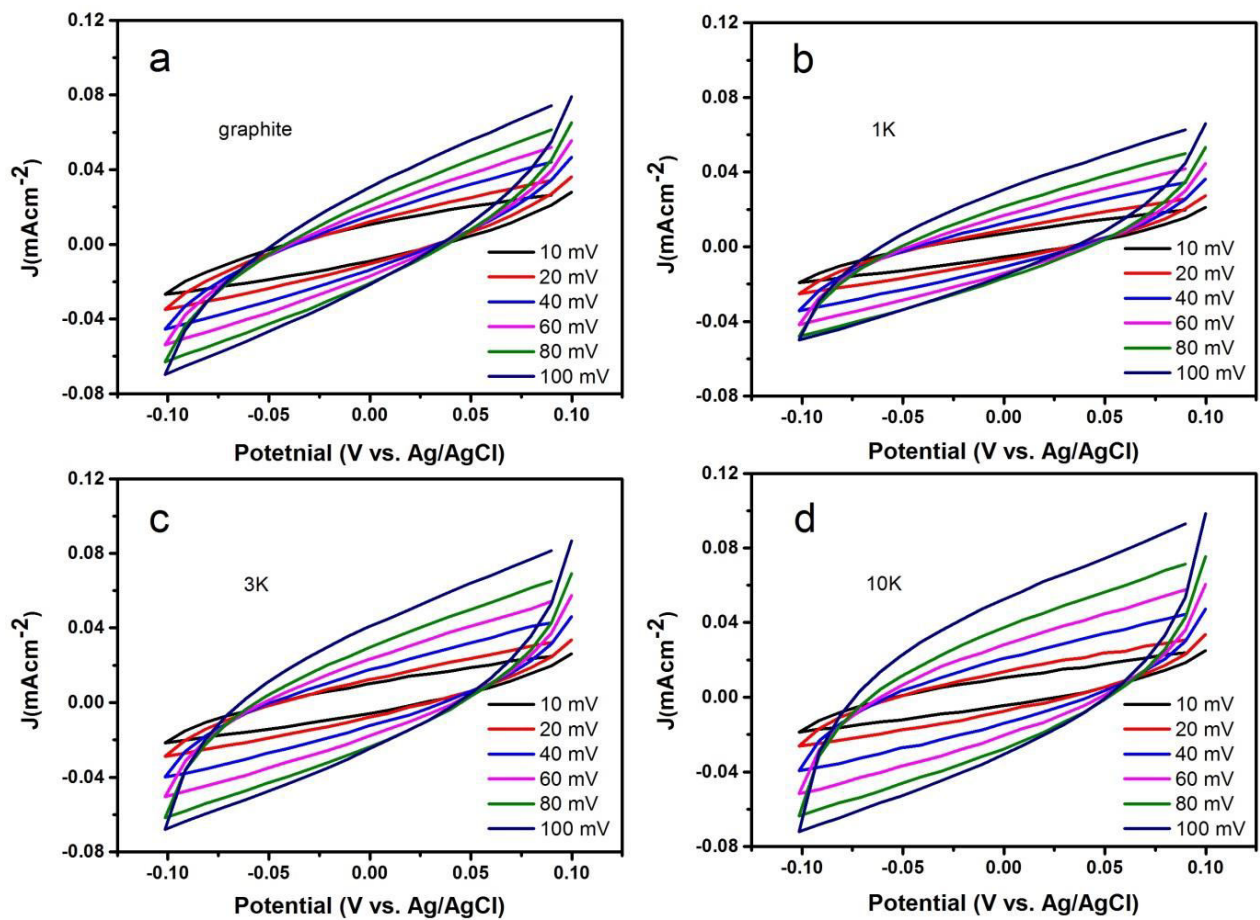

**Figure S18.** Cyclic voltammograms used to measure a non-faradic region between -0.1 to +0.1 V at various scan rates (0.1, 0.08, 0.06, 0.04, 0.02 and 0.01 Vs$^{-1}$).

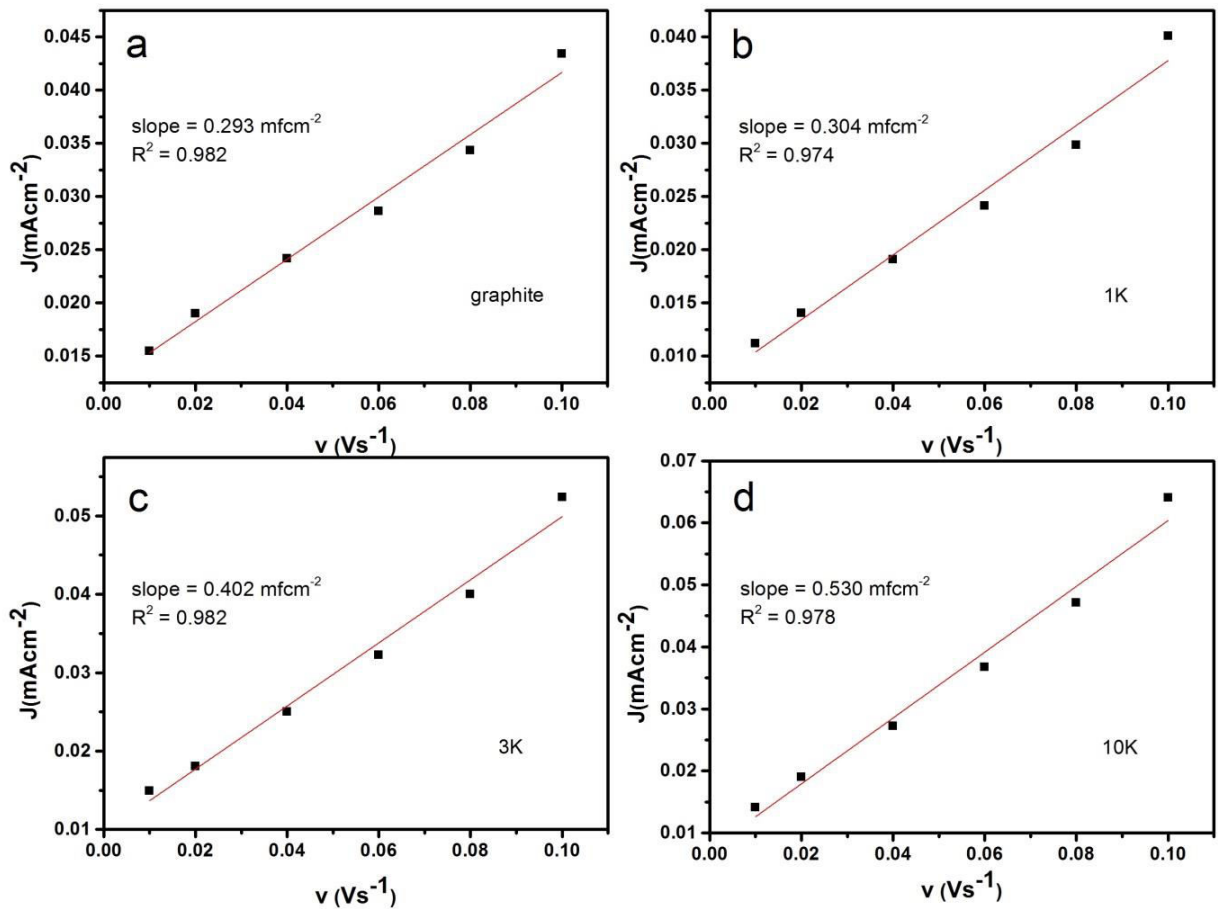

**Figure S19.** Subsequent capacitive current density measured at 0.025 V plotted as a function of scan rate. The average value of the slope was determined as the double-layer capacitance of each catalyst.

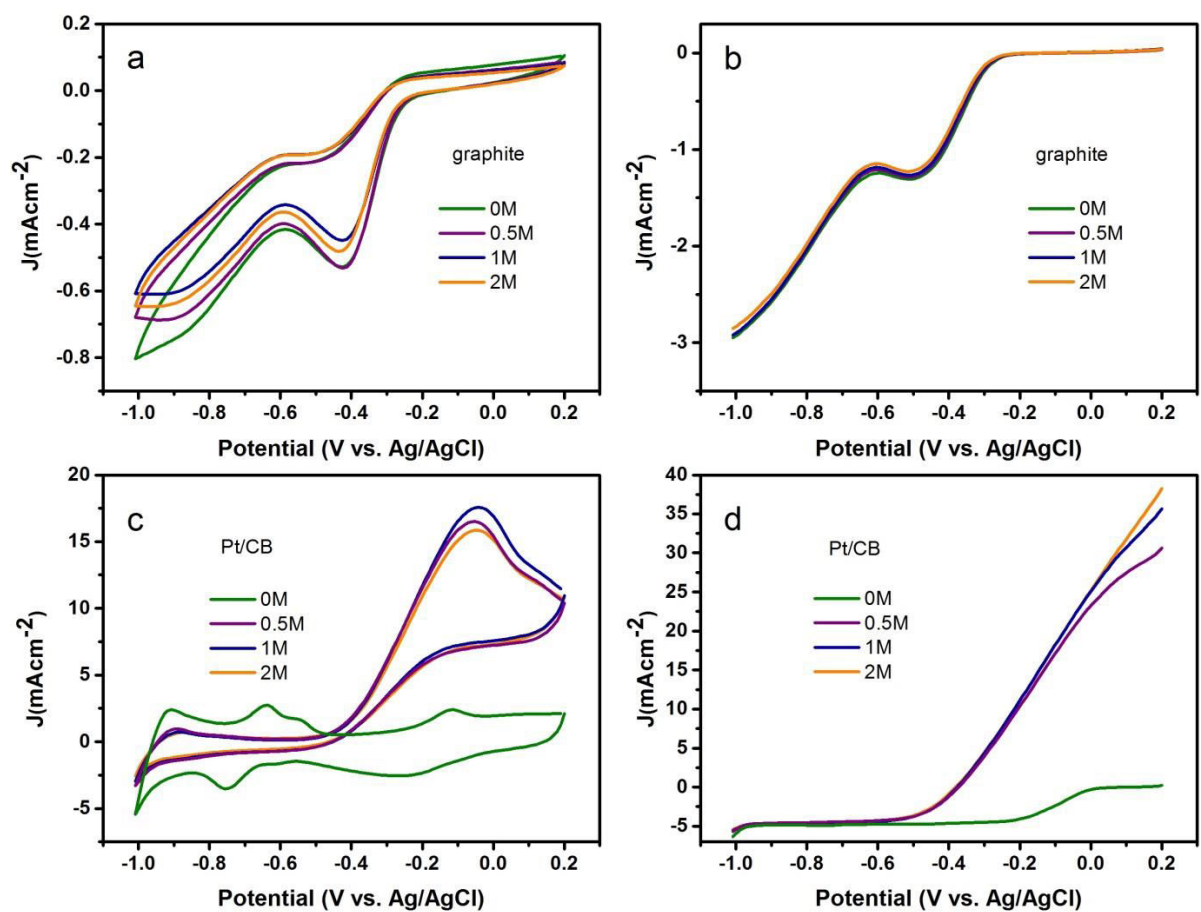

**Figure S20.** a) CV and b) LS responses of raw graphite after methanol addition. c) CV and d) LS responses of Pt/CB (20 wt % Pt) after methanol addition. All measurements were carried out in oxygen saturated 0.1M aq. KOH (3mm diameter GCE).

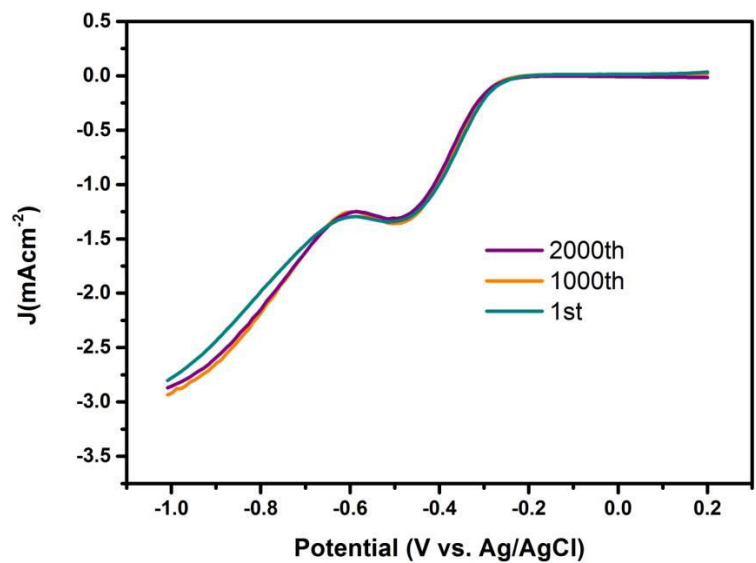

**Figure S21**. LS response of graphite after 1st, 1000th, and 2000th CV scan. Measurements performed in oxygen saturated 0.1M aq. KOH at a rotation speed of 1600 rpm (BASi, 3mm diameter GCE).

**Table S1.** XRD data of graphite, 1K, 3K, and 10K. Catalysts were dried under ambient conditions to form powder samples.

|  | **Peak Position (002)** | **D spacing (nm)** | **FWHM (002)** |
|---|---|---|---|
| **graphite** | 26.52° | 0.3358 | 0.243° |
| **1K** | 26.52° | 0.3358 | 0.257° |
| **3K** | 26.48° | 0.3363 | 0.408° |
| **10K** | 26.36° | 0.3378 | 0.502° |

**Table S2.** XPS data showing the compositions of graphite, 1K, 3K, and 10K. Catalysts were drop dried under ambient conditions on Si substrate.

|  | **C (at %)** | **O (at %)** |
|---|---|---|
| **graphite** | 98.92 | 1.08±0.28 |
| **1K** | 97.66 | 2.34±0.26 |
| **3K** | 97.92 | 2.08±0.50 |
| **10K** | 97.42 | 2.58±0.21 |

**Table S3**. XPS deconvolution data for C 1s.

| C 1s | C1 (284.47±0.04eV) % | C2 (285.40±0.06eV) % | C3 (287.21±0.02eV) % |
|---|---|---|---|
| **Graphite** | 83.83 | 16.17 | -- |
| **1K** | 78.89 | 18.20 | 2.91 |
| **3K** | 77.21 | 19.53 | 3.26 |
| **10K** | 75.60 | 20.67 | 3.73 |

**Table S4**. XPS deconvolution data for O 1s

| O 1s | O1 (530.77±0.10eV) % | O2 (532.00±0.08eV) % | O3 (533.11±0.29eV) % |
|---|---|---|---|
| **Graphite** | -- | -- | 100 |
| **1K** | 23.06 | 38.01 | 38.94 |
| **3K** | 24.19 | 39.76 | 36.05 |
| **10K** | 20.80 | 41.35 | 37.84 |

**Table S5.** Electrochemical activity of catalysts relative to pristine graphite based on exchange current densities.

| Sample | $j_0$ (Acm$^{-2}$) | Enhancement relative to graphite | Double layer Capacitance ($C_{DL}$) (mFcm$^{-2}$) | Surface are Relative to graphite | No. of active sits to surface area relative to graphite |
|---|---|---|---|---|---|
| Graphite | $8.4 \times 10^{-8}$ | 1.0 | 0.293 | 1.0 | 1.0 |
| 1K | $1.4 \times 10^{-7}$ | 1.7 | 0.304 | 1.04 | 1.6 |
| 3K | $3.7 \times 10^{-7}$ | 4.4 | 0.403 | 1.38 | 3.2 |
| 10K | $5.7 \times 10^{-7}$ | 6.7 | 0.530 | 1.81 | 3.7 |